# Finding optimal Noah-MP parameterizations for the characterization of surface heat fluxes in the Iberian Peninsula

David Donaire-Montaño[1] (0009-0007-8013-2650)

Matilde García-Valdecasas Ojeda[1,2] (0000-0001-9551-8328),

Nicolás Tacoronte[1] (0000-0001-7329-7931),

Juan José Rosa-Cánovas[1,2] (0000-0001-5320-3109),

Yolanda Castro-Díez[1,2] (0000-0002-2134-9119),

María Jesús Esteban-Parra[1,2] (0000-0003-1350-6150),

and,

Sonia Raquel Gámiz-Fortis[1,2] (0000-0002-6192-056X)

[1]Departamento de Física Aplicada, Universidad de Granada, Granada, Spain

dadonaire@ugr.es, mgvaldecasas@ugr.es, nicotacor@ugr.es, jjrc@ugr.es, ycastro@ugr.es, esteban@ugr.es, srgamiz@ugr.es

[2] Instituto Interuniversitario de Investigación del Sistema Tierra en Andalucía (IISTA-CEAMA), Granada, Spain

**Corresponding author:** Matilde García-Valdecasas Ojeda (mgvaldecasas@ugr.es)

**ABSTRACT**

Land surface models (LSMs) play a crucial role in the characterization of land-atmosphere interactions by providing boundary conditions to a regional climate model (RCM). This is particularly true over the Iberian Peninsula (IP), a region where a water-limited regime governs most of the territory. This work aims to optimize the Noah LSM with multiparameterization options (Noah-MP) configuration for characterizing heat fluxes in the IP when the Weather Research and Forecasting (WRF) model v3.9.1 is used as RCM. To do that, a set of 70 experiments with a 1-year length has been completed using 35 combinations of Noah-MP parameterizations, both for a year with dry conditions in the IP (2005 year) and for a year with wet conditions (2010 year). Land surface heat fluxes and soil moisture simulated with Noah-MP coupled to WRF (WRF/Noah-MP) have been evaluated using as reference the available FLUXNET station data and CERRA-Land reanalysis data. In general, the results indicate that WRF/Noah-MP accurately simulates soil moisture and surface heat fluxes over the IP, especially for wetter climate conditions. The clustering method has presented an optimal configuration from 10 groups (Clusters from A to J), which showed that the WRF/Noah-MP parameterizations with the greatest influence on the simulation of surface heat fluxes over the IP are canopy stomatal resistance (CRS), surface exchange coefficient for heat (SFC), soil moisture factor controlling stomatal resistance (BTR), runoff and groundwater (RUN), and surface resistance to evaporation/sublimation (RSF). In addition, dynamic vegetation (DVEG) seems to influence simulations. Although several clusters/configurations showed reasonable results, experiment s27I in Cluster I with Jarvis CRS, Chen97 SFC, CLM-Type BTR, BATS RUN, and Adjusted Sellers to decrease RSURF for wet soil for RSF seem to be more adequate to simulate surface heat fluxes in the IP.

**Keywords:** Noah land surface model with multiparameterization, Weather Research and Forecasting, surface heat fluxes, soil moisture, Iberian Peninsula.

# 1. Introduction

The land surface is a crucial component of the climate system. It integrates significant parts of the biosphere, lithosphere, cryosphere, hydrosphere, and continuously interacts with the atmosphere through the exchange of energy, mass, and momentum (Wallace and Hobbs, 2006). These interactions affect convection, the boundary layer, cloud formation, precipitation, floods, heat waves, wind, and other physical processes (Ardilouze et al., 2022; Cammalleri et al., 2017; Miralles et al., 2019, 2014; Seneviratne et al., 2006). Therefore, the land surface plays a significant role in weather and climate dynamics across different spatial and temporal scales.

Land surface models (LSMs), an essential component in regional climate models (RCMs), provide lower boundary conditions for representing the most relevant physical processes acting on the land surface. These processes include dynamic vegetation, stomatal resistance, runoff and groundwater flow, soil permeability, and albedo, among others. In this context, different LSMs have been adapted to RCMs. For the Weather Research and Forecasting (WRF, Skamarock et al., 2008) model, during the last decades, LSMs such as the Community Land Model (CLM, Lawrence et al., 2019), the Noah (Chen and Dudhia, 2001), and its latest developed version, the Noah LSM with multiparameterization options (Noah-MP, Niu et al., 2011) has been widely considered (Hu et al., 2023; Li et al., 2022; Ma, 2023; You et al., 2020; Zhang et al., 2016). The latter has been developed to be used under different environmental conditions worldwide, and for this reason it is more adaptable, allowing different parameterizations settings. Noah-MP uses mathematical equations to approximate key land processes by applying simplifications, such as setting equation parameters, using parameter tables, applying boundary conditions, implementing specific parameterizations, or defining initial conditions (Bonan et al., 2002; Chen and Dudhia, 2001; Liang et al., 1996; Niu et al., 2011). Consequently, the development of LSMs has also introduced uncertainties that need to be analyzed to improve the simulation of physical variables and their interactions in different contexts.

Noah-MP was incorporated as LSM in the WRF model from version 3.4 onwards. In these WRF versions, Noah-MP provides multiple options available for 12 key physical processes, allowing a more detailed representation of the physical processes involved in land-atmosphere interactions. Therefore, this LSM includes several parameterizations, each one with multiple options, resulting in many different combinations difficult to evaluate simultaneously (Li et al., 2022). Reducing these uncertainties in different contexts is both a significant challenge and a necessity. Consequently, numerous studies have been conducted across different regions worldwide, focusing on analyzing the effect of Noah-MP parameterizations in specific climate aspects (Li et al., 2022, globally; Zhang et al., 2021, 2016 in China and Tibet, respectively), dynamic vegetation influence (Li et al., 2022, globally; Yang et al., 2021, in China), energy and water exchange (Chang et al., 2020, in China; Li et al., 2022, globally; Ma, 2023, on the Tibetan Plateau), soil moisture representation (Cammalleri et al., 2017; Li et al., 2022, globally), snow-climate interactions (You et al., 2020, across global stations), and coupling strength (Zhang et al., 2021, 2022, in China, among others). In these processes, land-atmosphere energy fluxes play a critical role and are directly influenced by soil water content (Knist et al., 2017; Li et al., 2020; Miralles et al., 2019, 2014; Seneviratne et al., 2010, 2006). Therefore, studying surface heat fluxes and soil moisture and their relations could provide essential insights into weather and climate dynamics, climate feedback loops, enhance extreme weather prediction, and allow us to anticipate future climate conditions.

Regional climate simulations using WRF have been conducted over the Iberian Peninsula (IP) in recent years (Argüeso et al., 2012a, 2012b, 2011; García-Valdecasas Ojeda et al., 2021, 2020a, 2020b; Gómez-Navarro et al., 2010; Solano-Farias et al., 2024, among others). However, few studies have focused on gaining insights into the performance of LSM coupled to WRF. An example is the study by Jerez et al. (2010) where the ability of three LSMs (i.e., Noah, Pleim and Xu and Simple Five Layers) coupled to an RCM (i.e., MM5) in the characterization of IP temperature was analyzed. This study concluded that the use of an LSM generating more realistic surface fluxes is crucial, especially in a region such as the IP where there is a strong land-atmosphere coupling. Therefore, a deeper analysis of how LSMs, particularly Noah-MP, reproduce physical processes over an area with high complexity in terms of orography, climate, and land-cover types is necessary. In this context, this study aims to analyze the impact of using different parameterizations of Noah-MP coupled to WRF on the characterization of land-surface fluxes, i.e., sensible heat (SH) and latent heat (LH). This paper is structured as follows: Section 2 describes the data and methods applied, Section 3 details and discusses the results achieved, and Section 4 summarizes and concludes on the results of this study.

**2. Data and methods**

## 2.1. Configuration of the sensitivity experiments

WRF with the Advanced Research WRF dynamic core (WRF-ARW) version 3.9.1 has been used in this study to conduct WRF/Noah-MP sensitivity experiments. The selected WRF configuration is based on a two one-way nested-domain (Fig. 1a) approach. The coarser domain (d01) covers the European Coordinated Regional Climate Downscaling Experiment region (EURO-CORDEX, Jacob et al., 2014) with a spatial resolution of 0.44º (~50 km) on a 123 x 126 rotated latitude-longitude grid; and the nested domain (d02), is centered over the IP with 220 x 220 grid points at a horizontal resolution of 0.088º (~10 km, Fig. 1b). Vertically, the model extends up to 10 hPa, divided into 41 pressure levels. The physics schemes were: the Betts-Miller-Janjic (BMJ, Betts, 1986; Janjić, 1994) for cumulus; the WRF Single-Moment Three-Class (WSM3, Hong et al., 2004) for microphysics; the Asymmetric Convective Model version 2 (ACM2, Pleim, 2007) for the planetary boundary layer (PBL); and the Community Atmosphere Model 3.0 (CAM3, Collins et al., 2004) for radiation (both longwave and shortwave). This WRF configuration has previously demonstrated skillful performance in reproducing the main spatial patterns of primary climate variables such as precipitation and temperature in the IP (García-Valdecasas Ojeda et al., 2017). It has also shown good skill in simulating drought-related variables like soil moisture, evaporation (García-Valdecasas Ojeda et al., 2021, 2020a), and streamflow (Yeste et al., 2020). As land-cover map (Fig. 1c) and textures (Fig. 1d), the modified IGBP MODIS 20-category vegetation classification (Friedl et al., 2010) and the hybrid STATSGO/FAO 16-classes categories (Miller and White, 1998), respectively, have been used.

As initial and lateral boundary conditions (LBCs), which were updated at 6-hourly intervals, the fifth-generation reanalysis data from the European Centre for Medium-Range Weather Forecasts were used (ERA5, Hersbach et al., 2020). ERA5 has a horizontal resolution of 0.25º and provides data across 37 vertical levels from 1000 hPa to 1 hPa.

To analyze the effect of WRF/Noah-MP on the representation of heat fluxes and soil moisture, a set of 35 configurations (Fig. 2) were selected by combining different options of 11 Noah-MP parameterizations (Table 1). These have been set according to previous studies (Chang et al., 2020; Gómez et al., 2021; Li et al., 2022; Ma, 2023; Torres-Rojas et al., 2022; Zhang et al., 2020, 2022), while the remaining ones available in WRF v3.9.1 were set as the default values. These combinations have been used to carry out a total of 70 1-year simulations under different climate conditions (dry year and wet year) in order to analyze the effect of Noah-MP configuration under different water availability conditions. For this end, 2005 has been selected as dry year in the IP (Fig. 3), characterized by a deep drought and the occurrence of important heat wave events. On the other hand, 2010 has been selected as a very wet year in the IP (Fig. 3), with a large amount of precipitation across the IP.

Due to soil processes require a long spin-up period to reach their equilibrium and avoid inaccuracies in the initial soil moisture conditions (Jerez et al., 2020; Khodayar et al., 2015), spin up runs were completed to be used for all experiments with the same climatic conditions. Thus, for climate experiments, in dry and wet conditions independently, a 30-year spin-up run was completed using WRF/Noah-MP with default Noah-MP parameterizations, according to the methodology proposed by Hu et al. (2023). These sufficiently long simulations have been used for soil initial conditions, for each 1-year simulation separately (i.e., from 1975 to 2004 for the dry 1-year simulation and for 1980 to 2009 for the wet 1-year simulation).

### 2.2. Reference data

To compare the WRF/Noah-MP experiments, data from FLUXNET stations located in the IP (Fig. 1b), with available data for the years 2005 and 2010, have been also considered in this study. These data, provided by the European Fluxes Database Cluster (EFDC, https://www.europe-fluxdata.eu), include observations from eddy covariance flux tower stations, which were preprocessed, quality-checked, and corrected for instrumental errors. Among the stations located in the IP, those with data available for both years of study (2005 and 2010) were selected for comparison with WRF/Noah-MP experiments. The first station, Las Majadas de Tiétar (LMa), is situated in the central-western part of the IP (39.94°N, 5.77°W), at an altitude of 265 m. This station records a mean annual temperature of approximately 18.5°C and an average annual precipitation of 572 mm, falling mainly from November to March, with dry summers (Perez-Priego et al., 2017). It represents a tree-grass savannah ecosystem, predominantly composed of an herbaceous layer and scattered evergreen broadleaf oak trees. The nearest grid point in the WRF/Noah-MP model is classified as savannah. The second station, Llano de los Juanes (LJu), is located in the southeastern IP (36.93°N, 2.75°W), at an altitude of 1600 m. This site has a mean annual temperature of around 16°C and an average annual precipitation of approximately 400 mm, mostly falling during autumn and winter, with very dry summers. It is situated within a shrubland ecosystem (Serrano-Ortiz et al., 2007), coinciding with the same land use of its nearest grid point in WRF/Noah-MP grid. The third station, El Saler 2 (ES2), is located on the eastern coast of the IP (39.27°N, 0.32°W), with a mean annual temperature of about 18°C and an average annual precipitation of 550 mm. This station is situated within a rice paddy, which is flooded most of the year, even during the summer months (González-Zamora et al., 2019). The nearest grid point, classified as cropland, could be not adequate for comparison. This discrepancy could lead to significant differences in the simulation of heat fluxes due to differing soil moisture conditions. Nevertheless, including this station is valuable, as it allows the identification of potential limitations in WRF/Noah-MP and CERRA-Land in capturing local variability. Therefore, data from this station should be interpreted carefully. From these FLUXNET stations, we used the SH and LH outputs from flux measurements obtained from an eddy covariance tower and the soil water content (SWC) in percentage from complementary sensors, all of them with data collected at 30-minute intervals for both years, 2005 and 2010.

Additionally, SH, LH, and the volumetric soil moisture ($m^3/m^3$), from the Copernicus European Regional ReAnalysis for Land (CERRA-Land, Schimanke et al., 2021) have been also used as reference datasets. CERRA-Land provides near-surface atmospheric and soil fields every 3 hours at a horizontal resolution of 5.5 km from 1984 to the present. Previous studies have demonstrated that CERRA-Land reanalysis improves upon the global ERA5 for temperature, with this result being particularly clear for areas with complex terrain as the IP (Ridal et al., 2024). For comparison with the WRF/Noah-MP experiments, these data have been regridded to the WRF mesh using bilinear interpolation. Since SWC is expressed in percentage, and volumetric soil moisture in $m^3/m^3$, standardized bias in both variables were calculated by subtracting the experiment mean and dividing by the standard deviation, thus making the analysis comparable in terms of soil moisture (SMOIS). To ensure comparability between WRF/Noah-MP and CERRA-Land SMOIS data, the maximum common soil depth available in both datasets (2 m) was selected. Thus, SMOIS was integrated across the entire 2 m column for both datasets by summing the contributions from each soil layer: four layers in WRF/Noah-MP (0–0.10 m, 0.10–0.40 m, 0.40–1 m, and 1–2 m), and ten layers in CERRA-Land (0–0.01 m, 0.01–0.04 m, 0.04–0.10 m, 0.10–0.20 m, 0.20–0.40 m, 0.40–0.60 m, 0.60–0.80 m, 0.80–1 m, 1–1.5 m, and 1.5–2 m).

### 2.3. Sensitivity analysis of WRF/Noah-MP

#### 2.3.1. Experiments grouping

Since we are interested in identifying the Noah-MP parameterizations and their specific options playing a more significant role in improving the spatial and temporal performance of surface energy fluxes, a K-means clustering procedure has been applied to the total number of experiments.

The main overall aim of clustering techniques is to separate a dataset into groups so that comparable data appears in the same group but still exhibiting unique behavior with respect to elements present in other groups. To achieve this, the K-means clustering generates K groups using an optimization function such that the internal variability inside groups is minimized while the separation between them is maximized (Pampuch et al., 2023). It assigns elements to a cluster based on the smallest Euclidean distance to the centroid, which is the average of the cluster vector. This process is repeated iteratively until the centroids become stable. In our case, experiments with similar spatiotemporal patterns in surface energy fluxes were grouped. The model outputs were structured in a 35-row matrix (one per experiment), with 9038860 columns combining data from two years (2005 and 2010), two variables (LH and SH), and the 6191 grid points for 365 days per year.

Due to the high-dimensional nature of the data (i.e., two variables for two years in all grid points in the IP), the dimensionality was initially reduced using principal component analysis (PCA, Preisendorfer, 1988). This step also helps to filter out noise and redundant variability within the simulations, allowing the clustering algorithm to focus on the most relevant patterns (Wilks, 2006). Principal components (PCs) explaining over 90% of the variance were selected to feed the K-means algorithm, implemented using scikit-learn. Initial centroids were derived from an empirical probability distribution based on dataset inertia (sum of squared distances to the nearest centroid) and updated iteratively until clusters stabilized. We used 100 random initializations (n_init parameter) and a maximum iteration of 400 (max_iter). The optimal number of clusters was determined by maximizing the silhouette coefficient (Rousseeuw, 1987), a widely-used metric that evaluates how well-defined each cluster is.

#### 2.3.2. Performance of heat fluxes

SMOIS has a significant role in controlling heat fluxes from the soil (Achugbu et al., 2020; Klein et al., 2017; Knist et al., 2017; Seneviratne et al., 2010). Therefore, assessing the model's ability to reproduce this variable is crucial. Spring mean biases (considered as March-April-May, MAM) of SMOIS for both the dry (i.e., 2005) and wet (i.e., 2010) years, and root mean square error (RMSE), have been computed for every grid point to evaluate how well each experiment captures soil moisture. Spring has been selected because this season typically receives substantial solar energy and precipitation in the IP. Moreover, to further explore if the simulated energy fluxes time series reproduce the temporal variability from CERRA-Land, the Pearson correlation coefficients between the simulated and reanalysis data, previously removing the annual cycle, were computed (Kavvas and Delleur, 1975). Annual cycles of monthly SMOIS, SH, and LH from FLUXNET stations were also compared with those from WRF/Noah-MP experiments at the nearest grid point for each station. This validation was done because, although CERRA-Land provides a reliable reference for model evaluation (Ridal et al., 2024), it is a reanalysis product and not observations, and it is therefore recommended for assessing overall model behavior.

On the other hand, as another way to compare the WRF/Noah-MP outputs from different parameterization combinations, the impact of each experiment on the representation of the land-atmosphere coupling in the IP has been also analyzed. A useful way to visualize this coupling is by examining the correlation between SH and LH (García-Valdecasas Ojeda et al., 2020b; Knist et al., 2017; Seneviratne et al., 2010). This metric not only provides insight into how variables are coupled but also explains how radiative and soil moisture conditions influence land-atmosphere interactions. For example, the soil energy balance can be evaluated through the SH-LH coupling, as both variables compete for the available energy, which is regulated by soil moisture. In this context, SH-LH correlations help identify regions that are either energy-limited or water-limited. In energy-limited regions, soil moisture is sufficiently abundant to regulate surface temperature and, consequently, the near-surface atmospheric temperature. This results in simultaneous variations of SH and LH, showing a weak coupling between them. On the other hand, in water-limited regions, the lack of soil moisture in a region with enough energy constrains evapotranspiration (i.e., LH), which reduces near-surface atmospheric humidity and its gradient. As a result, near-surface temperature is primarily controlled by changes in soil temperature and, consequently, by SH flux. In this scenario, SH and LH are negatively correlated, leading to a strong land-atmosphere coupling. To evaluate how the experiments represent the spatial pattern of land-atmosphere feedback processes, the SH-LH coupling metric is assessed through Pearson correlation coefficients of the different experiments and CERRA-Land for both the dry (i.e., 2005) and the wet (i.e., 2010) years.

**2.3.3. Final determination of optimal experiments**

Finally, the analysis proposed by Li et al. (2022) was conducted to differentiate the performance and determine the optimal configuration for characterizing surface heat fluxes in the IP. The non-parametric Kling-Gupta Efficiency (KGE) metric (Pool et al., 2018) was calculated as a skill score metric from daily values. The KGE captures the variability, bias, and dynamics of the temporal series for each grid point, following Equation 1.

$$KGE = 1 - \sqrt{(\beta - 1)^2 + (\alpha - 1)^2 + (r_S - 1)^2} \quad (1)$$

Where $\beta$ is the mean bias parameter, computed as the ratio of the simulated and reference mean heat fluxes ($\overline{F_{exp}(t)}$ and $\overline{F_{ref}(t)}$, respectively) (Equation 2). $\alpha$ is the variability parameter, computed using Equation 5, which uses the Flow Duration Curve (FDC) from Equation 4 for both experimental ($FDC_{exp}$) and reference ($FDC_{ref}$) data. In Equation 3, $F_{sorted}$ represents the temporally sorted flux series, and in Equation 4 $n$ is the number of elements in the time series. Finally, $r_S$ in Equation 1 is the temporal Spearman correlation coefficient between the experimental and reference data. KGE ranges from $-\infty$ to 1, with a perfect fit being represented by a value of 1.

$$\beta = \frac{\overline{F_{exp}(t)}}{\overline{F_{ref}(t)}} \quad (2)$$

$$F_{sorted} = sort(F(t)) \quad (3)$$

$$FDC = \frac{F_{sorted}}{n \cdot \overline{F(t)}} \quad (4)$$

$$\alpha = 1 - \frac{1}{2} \sum_{i=1}^{n} |FDC_{exp}(i) - FDC_{ref}(i)| \quad (5)$$

In this context, the non-parametric KGE was calculated for daily SH and daily LH data for 2005 and 2010 separately, and these values have been used to rank experiment performance for both variables and years. Then, the highest rank was used to select the best cluster for each grid point, variable, and year. Additionally, to determine the variability of the best cluster, the amplitude of the ranges was calculated for each grid point.

### 3. Results and discussion

#### 3.1. Experiments grouping

A K-means clustering was applied to group experiments with similar behavior. Previously, PCA was applied to reduce redundancy, transforming the 35 x 9038860 matrix into a 35 x 17 matrix, as 17 PCs captured over 90% of the variance. To select the optimal number of clusters, the silhouette coefficient was used, which reached a maximum in 10. Table 2 summarizes how experiments are distributed among groups and the different parameterization options. Detailed information on the experiments and their inclusion in the clusters also is provided in Fig. 2.

Cluster A is composed of 4 experiments (s0A, s6A, s21A, and s30A), which have in common 5 parameterization options (all default options for CRS, SFC, FRZ, RAD, and RSF). Cluster B (s1B, s5B, s7B, s8B, s9B, and s11B) includes seven parameterization options in common (CRS, SFC, BTR, INF, TBOT, and RSF). Experiments composing Cluster C (s2C and s22C) have all the options in common except the DVEG. Experiments from Cluster D (s3D, s10D, s13D, s17D, s18D, and s25D) have in common CRS, SFC, FRZ, RAD, and RSF. However, for this cluster the BTR is different from Noah, and all show dynamic vegetation. Cluster E (s4E, s16E, and s32E) is characterized by those combinations with the SFC parameterization set to the original Noah (Chen97), with this one of the few clusters presenting this parameterization option. Cluster F (s12F, s14F, and s15F) is composed of experiments with seven parameterizations in common, as they have the same DVEG, CRS, SFC, BTR, RAD, ALB, TBOT, and RSF. In Cluster G there are five experiments (s19G, s20G, s23G, s24G, and s29G) sharing five common parameterizations (DVEG, CRS, SFC, RAD, and ALB), and all of them show a BTR different from Noah. Cluster I (s27I, s28I, and s31I) is composed of all experiments where Jarvis CRS option is used, suggesting that CRS has a significant impact on the configuration. Finally, Clusters H (s26H) and J (s33J and s34J) have in common the RUN option TOPMODEL with equilibrium water table (EQWT).

Furthermore, comparing pairs of experiments, additional conclusions can be drawn. If two experiments share all parameterizations except one but are in different clusters, it suggests that WRF/Noah-MP is sensitive to that parameterization. Conversely, if they belong to the same cluster, it indicates a smaller effect. In this regard, the comparison of s1A vs. s4E shows that WRF/Noah-MP is sensitive to SFC options, with s1A and s4E (s32E) using M-O and Chen97, respectively. Similarly, the comparison of s1A and s2C indicates that dynamic vegetation (On-Dickinson vs. OFF-Calc) affects the WRF/Noah-MP performance, but the comparison between s2C and s22C indicates that the way to calculate GVF seems not to be so important, at least when LAI is calculated with a look-up table and GVF is derived from the shadow fraction. Moreover, for RSF, comparisons like s11B/s12F or s13D/s14F show that WRF/Noah-MP is sensitive to the way this parameterization is modeled, at least when SZ19 and AS-WET are compared. The comparisons between s1A/s3D, s27I/s28I, or s1A/s10D, however, suggest sensitivity to BTR, particularly when comparing Noah with CLM (s1A/s3D, s27I/s28I) or SSiB (s1A/s10D).

Interestingly, s1B, s5B, and s6A have all parameterization options in common except RAD. This discrepancy could be because s1A and s5B, and s6A use ModTS, GridTS, VegTS, respectively, and while ModTS and GridTS are two simplified schemes that are similar, VegTS is more complex, as it considers the separation between vegetation and soil fractions (Niu and Yang, 2004). Similarly, the comparison between s1A vs. s8B (s19G vs. s24G) seems to indicate that the GW and BATS RUN (GW and SR) options have, in general, similar behavior. However, as mentioned above, combinations with EQWT seem to have a large effect in separating the experiments in groups. In contrast, the comparisons s1A vs. s7B, s1A vs. s9B, and s19G vs. s20G suggest that ALB, FRZ, and TBOT have not clear effects.

### 3.2. Soil moisture and its effect on land-surface fluxes

Fig. 4 shows the bias in terms of SMOIS for all the experiments compared to CERRA-Land for the spring season in 2010. Similarly, Fig. 5 shows the bias in terms of SH (Fig. 5a) and LH (Fig. 5b) in W/m$^2$, for the same season. In both figures, positive biases represent overestimations in relation to CERRA-Land, and negative biases represent underestimations. The mean bias for SMOIS, SH, and LH for the different experiments in 2005 (dry year), along with RMSE spatial patterns, are provided in the supplementary material (Figs. S1 to S5).

Two main spatial patterns in SMOIS bias were observed when comparing WRF/Noah experiments to CERRA-Land data. Experiments from Clusters H (s26H) and J (s33J and s34J) exhibited an overestimation in the north-northwest of the IP, high altitude regions, and part of the southern region, and strong negative biases in large semi-arid areas (central, southern, and eastern IP). Both clusters use the EQWT RUN parameterization, a simplified hydrological model suitable for global-scale simulations (Niu et al., 2011). This result may suggest that this RUN parameterization option could not be appropriate to simulate runoff over IP, at least at high spatial resolution. In addition, the largest biases in this case occur in semi-arid areas, showing that EQWT option for RUN behaves inadequately, especially over water limited regions. However, the results differ in terms of SH (Fig. 5a), with Cluster J showing a stronger positive bias than Cluster H, but both show similar behavior for LH (Fig. 5b). These clusters, among other parameterizations, differ in SFC (i.e., Cluster J utilizes M-O, whereas Cluster H uses Chen97), parameterization that modulates surface processes and that shows a significant influence for SH modulation, with a lower overestimation of SH found when using Chen97 (in s26H). During the dry year, experiments from Clusters J and H also show a different behavior to the others in terms of SMOIS bias and RMSE (Figs. S1 and S2a respectively, in supplementary), which is extended from the southeastern to a large part of the north and northwestern IP. This leads to a generalized overestimation of SH for Cluster J and underestimation for Cluster H in the southern IP (Fig. S3a in supplementary material). However, LH is overall underestimated (Fig. S3b in supplementary material) for both clusters.

The remaining experiments mainly show widespread SMOIS overestimations compared to CERRA-Land for 2010 (Fig. 4), which are more pronounced depending on the cluster. The most pronounced overestimations appear in s0A, s30A, s25D, s23G, S24G, which are more marked in the southern half of the IP. For 2005, However, experiments from Clusters A, B, C or I show similar patterns with overall overestimation while other experiments show also underestimations mainly over the south (Fig. S1).

In the comparison between clusters, Clusters B (s1B, s5B, s7B, s8B, s9B, s11B) and C (s2C and s22C) exhibit similar behaviors, with a generalized overestimation of SMOIS during spring 2010 (Fig. 4), slightly stronger in Cluster C. This pattern is also observed in 2005 (Fig. S1), but with more pronounced positive biases. In terms of heat fluxes, LH and SH are overestimated in Cluster B (Fig. 5a, 5b), while Cluster C exhibits a strong SH overestimation (Fig. 5a) and slight LH underestimation (Fig. 5b). The key difference between these clusters is LAI calculation; Cluster B uses the ON option, while Cluster C uses OFF.

The experiments from Clusters D and G also exhibit similar behavior. On the one hand, experiments s23G, s24G, and s25D exhibit pronounced positive biases in term of SMOIS in a large part of the IP (Fig. 4), all using the SR RUN option (Fig. 2). In contrast, remaining experiments from these clusters (s3D, s10D, s17D, s18D, s19G, s20G, and s29G) show a very good agreement with CERRA-Land and they use either the GW or the BATS RUN options. In terms of the surface heat partitioning (Fig. 5), all experiments in Clusters D and G also show a similar pattern, with under- and overestimations in terms of SH (Fig. 5a) and a generalized overestimation in terms of LH (Fig. 5b), the latter being stronger for s23G, s24G and s25D. For 2005 (Figs. S1 and S3 in the supplementary material), however, experiments from both clusters behave similarly, showing all similar agreement with CERRA-Land for both SMOIS and heat fluxes, which could be suggesting that WRF/Noah-MP is more sensitive to RUN options in the wet year. In a similar way, experiments in Cluster F (s12F, s14F and s15F) show slight overestimations in SMOIS in spring 2010, less marked for s14F and s15F (Fig. 4). For 2005 s12F shows marked overestimations in the north while s14F and s15F indicate high underestimations in the southern half and over the northeast. These differences are mainly due to the BTR options, as s12F uses Noah, while s14F and S15F use CLM. However, in terms of heat fluxes, all experiments show similar behavior (Fig. 5), with generalized SH and LH overestimations. In spring 2005, SMOIS biases are more pronounced (Fig. S1), but biases are less evident for heat fluxes (Fig. S3).

Clusters I (s27I, s28I and s31I) and E (s4E, s16E and s32E) show SMOIS overestimations, but also some underestimations in both years, with Cluster E showing better performance, at least in terms of SMOIS (Fig. 4 and Fig. S1) Cluster E tends to overestimate LH (Fig. 5b), while Cluster I shows weaker bias (Fig. 5b). In terms of SH (Fig. 5a), however, both show a slight underestimation, which is a distinguishing feature of these clusters. These clusters share the Chen97 SFC, suggesting its role in SH underestimation, also corroborated during the dry year (Fig. S3).

Fig. 6a presents the temporal correlation (r) for simulated SH and CERRA-Land in the dry year (2005). A consistent pattern is observed across most simulations, with higher correlation values in the northwestern (r > 0.6), where large areas with savannah and wet savanna grid-points in WRF are located (Fig. 1). In contrast, the poorer correlation appears in the northeast, in areas mainly covered by mixed forest and open shrubland. However, although land use appears to have some influence, other factors could be also affecting the performance of heat flux simulations, so any association with land use should be interpreted with caution. Additionally, differences are also evident between experiments. Clusters H and J exhibit the lowest correlation values, followed by Clusters E and I. This fact could be attributed to the SFC parameterization, as Chen97 is used in Clusters H, E, and I. In fact, comparing Clusters H with J, the presence of Chen97 in H aligns with the worst correlation in terms of SH (Fig. 6, s26H). For the remaining experiments, differences are less clear, although some experiments such as s15F or s13D seem to show a better agreement with CERRA-Land in terms of SH.

In terms of LH (Fig. 6b), correlations are lower overall, with r values below 0.50 in a large part of the IP. LH is relatively well represented in the northwestern and western regions, predominantly covered by wet savanna and savanna, as occurred for SH. However, performance decreases in areas where woody savanna type are widespread, with some experiments displaying particularly low correlation values in these areas. The results also show that Clusters H and J consistently present a very poor performance, with r values below 0.2 in the whole IP. In contrast, Cluster I exhibit the best overall agreement with r values above 0.4. The second-best performing cluster is Cluster E, which shares with Cluster I the Chen97 SFC option. Additionally, Cluster F seems to be also good when we focus on the western half of the IP. For 2010, similar conclusions can be drawn regarding the temporal correlations of SH and LH (Fig. S6 in the supplementary material), which generally show higher correlation coefficients, indicating a better agreement with CERRA-Land.

The previous results evaluate the simulated SMOIS and surface heat fluxes for all the experiments in relation to CERRA-Land. However, as far as possible, they should be corroborated by comparison with observations. For this reason, SH, LH, and SMOIS annual cycles have been compared with those from stations at three specific FLUXNET locations in both the dry (2005, Fig. 7a) and the wet (2010, Fig. 7b) years. Both years show an inverse relationship between SH and LH during central months: LH reaches its maximum value during the spring season (MAM), while it drops to minimum values during the summer (June-July-August, JJA). In contrast, SH reaches its peak in summer. This is a consequence of the differences in soil water availability between spring and summer, combined with the higher soil energy availability during spring and summer (Knist et al., 2017; Seneviratne et al., 2010). This behavior is consistently observed at all three station locations, which are situated in a transitional region where soil moisture availability strongly regulates the land-atmosphere coupling. In these areas, variations in soil water content control the partitioning of energy into latent and sensible heat, explaining the observed inverse relationship between SH and LH (García-Valdecasas Ojeda et al., 2020b). Overall, the results show that WRF/Noah-MP experiments capture the SMOIS and surface heat fluxes annual cycles, better during the wet year, although with a generalized overestimation. As an exception, the ES2 station presents marked differences between observations, CERRA-Land and WRF/Noah experiments, in terms of heat fluxes and SMOIS. Additionally, for this location, CERRA-Land and WRF/Noah-MP do not show the same partitioning of the heat fluxes as observations from April onwards. This is attributed to the specific local conditions of this station, which are not adequately represented by WRF/Noah-MP and CERRA-Land. The station is located in an area that remains flooded for extended periods (González-Zamora et al., 2019), which explains the higher LH and lower SH values observed in the station data, especially during the central months of the year (Fig. 7). Concerning the differences between experiments, in general, the results evidence that Cluster J presents problems to correctly simulate the annual cycle of SMOIS and subsequently the LH. This is consistent with the findings in Figs. 4, 5, and 6, and could be attributed to issues with the functioning of the RUN EQWT option. Although no cluster shows a clearly better result, Clusters E and I, followed by Cluster G tend to present more similar values to CERRA-Land and observations, at least in terms of intra-annual variability.

### 3.3. Land-atmosphere coupling analysis

Fig. 8 shows SH-LH Pearson correlations coefficients for the years 2005 and 2010, as a measure of land-atmosphere coupling. Positive SH-LH correlations indicate atmospheric control without land-atmospheric coupling (i.e., the absence of latent flux exchange is due to the lack of radiation), while correlations close to -1 mean that the soil impacts on the atmosphere, and thus, there is a strong land-atmosphere coupling. These SH-LH correlations are shown for all WRF/Noah-MP experiments and CERRA-Land, used as a reference. Additionally, the pattern correlation (r) and spatial root mean squared error (RMSE) are displayed in the bottom right corner of each figure in the panel.

In CERRA-Land, the transitional zone (i.e., region with strong land-atmosphere coupling and therefore arid conditions) is well defined between the northern and mountain regions, where strong SH-LH positive correlations, typical of wet areas, are shown. In general, WRF/Noah-MP experiments present more difficulties capturing SH-LH correlations in the dry year, likely due to the SMOIS overestimations. Concerning differences between experiments for the dry year (Fig. 8a), the results show that Cluster C exhibits difficulties in characterizing the coupling strength over the IP, showing positive correlation in most of the region. For this cluster the pattern correlation is around 0 and the spatial RMSE with respect to CERRA-Land reaches values close to 0.8. The latter also occurs for Cluster A, at least in s0A and s6A, and for Cluster B (r close to 0.5 and spatial RMSE around 0.6). Clusters H and J also show inadequate behavior, presenting an absence of correlations throughout the region. Cluster D can detect some coupling but underestimates it, with a spatial RMSE of up to 0.30 and pattern correlation below 0.8. Cluster I has an intermediate behavior, and Clusters G, F, and E perform the best overall. For 2010 (Fig. 8b), all experiments seem to show better agreement with CERRA-Land with higher pattern correlation and lower spatial RMSE. This behavior is especially shown in Clusters A, B, C, and G.

Comparing pairs of experiments, we can also draw some conclusions in terms of SH-LH coupling. For example, s1B vs s4E shows Chen97 outperforms M-O in representing land-atmospheric coupling, especially in the dry year (Fig. 8a). The comparison of s1B vs. s2C/s22C suggests dynamic vegetation improve coupling. Additionally, s1B vs. s3D or s27I vs. s28I highlight that CLM BTR better represents the coupling than Noah BTR. However, in terms of RUN, only experiments with EQWT (s26H, s33J and s34J) present significant differences with the remaining options.

### 3.4. Determination of the optimal experimental set

Fig. 9 shows the spatial distribution of the best-performing clusters for both years (2005 and 2010) and for both variables, SH and LH, separately. Additionally, Table 3 presents the percentage of area where each cluster is considered the best option, along with its relative amplitude. For both years and variables, the results show that Cluster I outperforms other clusters in terms of KGE, showing a higher percentage of area, especially for the LH variable (37.30% and 35.40% of area for 2010 and 2005, respectively). For SH, Cluster I also shows the best results (20.44% and 30.60% for 2005 and 2010, respectively), but is followed closely by other clusters. This is the case of Cluster F, especially for the dry year (2005) showing a 20.12% and 24.01% of area for SH and LH, respectively, although the area percentage of this Cluster is lower for SH in 2010. In this case (SH in wet year), Cluster G could be a better option than Cluster F with 17.28% of the associated area.

Therefore, since different parameterizations combinations cannot be selected for each grid point, it is appropriate to select the cluster with the best performance for the entire IP. In general, Clusters I, F, E and G seem to be the clusters presenting the best results in terms of KGE. However, Cluster I outperforms the others across much of the eastern half of the IP, for both SH and LH under wet conditions (year 2010). In the dry year (2005), Cluster I maintains its dominance in the eastern region for LH, but its performance for SH is disrupted by the influence of other clusters, such as Cluster F and G. In contrast, in the western half of the IP, the Cluster I performance ranking is less clear, especially for SH, with other clusters such as F and G gaining percentage of area associated. Note that Cluster I presents difficulties representing SH during the dry year (Fig. S3a in the supplementary material). This could be explained by the difficulties that Jarvis CRS face when drought conditions are prolonged (Qi et al., 2023). Concerning the amplitude between experiments (Table 3), Cluster I also presents overall lower percentages, ranging from 18.16% for SH in 2010 to 29.35% for LH in 2005. Specifically, experiments s27I and s28I could be the most representative experiments within Cluster I, presenting higher coupling pattern correlations and lowest RMSE values with respect to CERRA-Land than s31I (Fig. 8). Experiments s27I and s28I share the Jarvis CRS with OFF DVEG, Chen97 SFC and BATS RUN, differing in BTR (s27I uses CLM vs. s28I that uses Noah).

## 4. Conclusions and Discussion

Based on a wide set of WRF/Noah-MP experiments, which combined several options of Noah-MP parameterizations (DVEG, CRS, SFC, BTR, RUN, FRZ, INF, RAD, ALB, TBOT, and RSF), a sensitivity analysis has been performed over the IP. The evaluation has been carried out using 1-year length simulations for two years with different climate conditions: 2005 characterized by dry conditions and 2010 as a wet year. Simulated surface heat fluxes and soil moisture in a 10-km spatial resolution WRF/Noah-MP have been compared with CERRA-Land reanalysis data and FLUXNET observations. The main findings can be summarized as follows:

1. Overall, WRF/Noah-MP is able to characterize surface heat fluxes over the IP. However, under dry conditions it shows more problems for characterizing them than for wet conditions. This result is evidenced by a better agreement with FLUXNET observations in terms of the annual cycle of monthly values in 2010, but also by higher correlations and lower bias compared to CERRA-Land.

2. The cluster analysis suggests that RUN, SFC, BTR, RSF, and CRS parameterizations have the greatest influence on surface heat fluxes and soil moisture over the IP. In addition, dynamic vegetation seems to influence simulations, at least as far as the calculation of LAI is concerned. In contrast, ALB, TBOT, FRZ, and INF do not show apparent differences between experiments, as also was shown by Zhang et al. (2016). These results agree with other studies (Chang et al., 2020; Gan et al., 2019; Gómez et al., 2021; Hosseini et al., 2022; Hu et al., 2023; Li et al., 2022; Zhang et al., 2020, 2016), who performed sensitivity analyses of Noah-MP worldwide.. In this work, despite the large number of possible combinations, these key parameterizations appear to be central to the differentiation between groups. The results also show different behavior depending on the options used in RAD and BTR, indicating that the complexity of the parameterizations is a key factor in grouping experiments. However, regional factors may influence these effects. For instance, You et al. (2020) highlighted the role of TBOT parameterization in snow climates, and Li et al. (2022) emphasized the importance of ALB parameterization in simulating snow depth through heat fluxes. Therefore, the effects of WRF/Noah-MP parameterizations are complex due to interactions in soil processes. Soil moisture significantly influences local weather and plant physiology through interconnected mechanisms (Bonan, 2008), affecting surface heat fluxes. In years with low soil moisture, evapotranspiration decreases, leading to a rise in sensible heat flux and, consequently, higher temperatures. Additionally, dry soils can limit stomatal opening, disrupting the carbon balance in ecosystems and affecting CRS parameterization. Furthermore, reduced soil moisture may enhance soil water retention by promoting infiltration and decreasing runoff (affecting RUN parameterization). This, in turn, could influence evapotranspiration rates (BTR parameterization) and impact vegetation growth (DVEG) (Cai et al., 2024). Chang et al. (2020)

and Neukam et al. (2016) have demonstrated as BTR and CRS parameterization are linked due to the dependence of canopy resistance on plant transpiration and stomatal resistance.

3. The RUN parameterization appears to play a relevant role in the ability of WRF/Noah-MP to characterize both surface heat fluxes and soil moisture in the IP. The EQWT, a simplified hydrological model developed for global-scale simulations (Niu et al., 2011) and used in Clusters H and J, produces notable biases compared to CERRA-Land and FLUXNET stations across all the analyses and under different climatic conditions. This is suggesting that EQWT is unsuitable for simulating runoff over the IP, at least at this spatial resolution. In addition, the largest biases in this case occur in semi-arid areas, so this result could be corroborating the results found by Zheng et al. (2019), who evidenced that RUN parameterization acquires great relevance in water-limited regions. Conversely, the BATS scheme, GW scheme, and SR schemes seem to be suitable options in the characterization of heat fluxes in the IP, with GW and BATS outperforming SR. This latter is especially shown in terms of SMOIS bias. These results partly agree with those of Chang et al. (2020), who found GW effectively represented baseflow runoff, while EQWT showed unrealistic precipitation response, and with Gan et al. (2019) who found BATS as optimal for SH and SR for LH.

4. For SFC parameterization, experiments with Chen97 option result in lower SH compared to those with M-O, finally resulting in underestimations in this variable compared to CERRA-Land. This behavior is more notable in the dry year. However, for the other variables, it produces comparable results to M-O. Zheng et al. (2019) pointed out that M-O leads to higher canopy evaporation than Chen97, which could be the explanation for the lower SH found when we used Chen97. While Chen97 considers the difference between the roughness length for heat and momentum, M-O considers the zero-displacement height. Yang et al. (2011) found that M-O was able to correct land skin temperature cold biases in arid western regions in the U.S. produced by Chen97. This correction was mainly attributed to the surface exchange coefficient, with M-O leading to smaller values than Chen97. Moreover, the M-O SFC option together with Jarvis and OFF DVEG parameterization seem to result in better correlations in terms of LH, but worse in terms of SH for the IP. WRF/Noah-MP experiments present more difficulties capturing land-atmosphere coupling in the dry year, likely due to the SMOIS overestimations. Cluster I shows the best agreement overall with CERRA-Land for this variable. These results are in agreement with Zhang et al. (2016), who assessed Noah-MP uncertainties in the Noah-MP in Tibet, founding that Jarvis (CRS option used in Cluster I) was better for simulating LH, while Ball-Berry (CRS option in Clusters F, G and E) was better for SH.

5. The best performance of the WRF/Noah-MP parameterization combinations vary along the IP. This result is consistent with other studies that demonstrated the model's sensitivity to parameterization scheme selections depends on the specific region analyzed, as simulation performances also vary with large-scale characteristics such as climatic conditions, land cover, soil textures, and geographical features (Hong et al., 2014; Zhang et al., 2022). Therefore, the choice of better configuration may be related to topography, land-cover and soil textures of the region, and as pointed out by other studies (Li et al., 2022, 2020; Zhang et al., 2016) Note that the influence of soil texture is directly related to BTR, as this affects soil moisture according to the texture type and directly controls plant transpiration (Niu et al., 2011). This effect is clearly seen in the bias of soil moisture, since experiments in Cluster F are more or less biased depending on the type of BTR used, with CLM outperforming Noah. Therefore, further research into this topic is needed in order to elucidate such effects.

6. For the whole IP climate simulations, Cluster I could be selected as the optimal one to characterize the surface energy fluxes, although there are other combinations that could be also good options (such as Clusters F, E, and G), with the exception of Clusters J and H. Within Cluster I, the experiment that provides the best representation of surface heat fluxes is s27I, which is configured with the following Noah-MP options: the CLM-type BTR, OFF-Calc for DVEG, BATS for RUN, Jarvis for CRS, Chen97 for SFC, NonLinear for INF, VegTS for RAD, CLASS for ALB, TBOT8m for TBOT, and AS-Wet for RSF. Some of these options for the optimal combination are in agreement with the results found from other studies. Chang et al. (2020), in their sensitivity study over a subtropical forest on China found that Chen97 option for SFC improves the M-O option and Jarvis improves the Ball-Berry option for CRS (which conduct to the use the OFF option for DVEG). This scheme seems to simulate more effectively land surface ventilation (Niu et al., 2011).

The results from this work show that the response of WRF/Noah-MP to simulate surface heat fluxes in the IP depends on the selected Noah parameterizations, one of the most evident being land surface states and processes, and also of the climatic conditions. The optimization of land surface models is challenging due to the model complexity, uncertainty along with the high computational cost involved. Although further research is necessary to analyze not only the underling physical processes and their complex interconnections, but also the impact from cover land use, soil textures and topography of the region, the optimal configuration selected in this work could be helpful in accurately characterizing land-atmosphere coupling and further climate simulations for the IP as a whole.

**Acknowledgments**

This work has been financed by the Project PID2021.126401OB.I00. funding by MICIU/AEI/ 10.13039/501100011033 and by FEDER, UE. Authors thank the anonymous referees for their insightful comments that have helped to improve this work.

# References

Achugbu, I.C., Dudhia, J., Olufayo, A.A., Balogun, I.A., Adefisan, E.A., Gbode, I.E., 2020. Assessment of WRF Land Surface Model Performance over West Africa. Advances in Meteorology 2020, 1–30. https://doi.org/10.1155/2020/6205308

Ardilouze, C., Materia, S., Batté, L., Benassi, M., Prodhomme, C., 2022. Precipitation response to extreme soil moisture conditions over the Mediterranean. Clim Dyn 58, 1927–1942. https://doi.org/10.1007/s00382-020-05519-5

Argüeso, D., Hidalgo-Muñoz, J.M., Gámiz-Fortis, S.R., Esteban-Parra, M.J., Castro-Díez, Y., 2012a. Evaluation of WRF Mean and Extreme Precipitation over Spain: Present Climate (1970–99). Journal of Climate 25, 4883–4897. https://doi.org/10.1175/JCLI-D-11-00276.1

Argüeso, D., Hidalgo-Muñoz, J.M., Gámiz-Fortis, S.R., Esteban-Parra, M.J., Castro-Díez, Y., 2012b. High-resolution projections of mean and extreme precipitation over Spain using the WRF model (2070–2099 versus 1970–1999). J. Geophys. Res. 117, 2011JD017399. https://doi.org/10.1029/2011JD017399

Argüeso, D., Hidalgo-Muñoz, J.M., Gámiz-Fortis, S.R., Esteban-Parra, M.J., Dudhia, J., Castro-Díez, Y., 2011. Evaluation of WRF Parameterizations for Climate Studies over Southern Spain Using a Multistep Regionalization. Journal of Climate 24, 5633–5651. https://doi.org/10.1175/JCLI-D-11-00073.1

Ball, J.T., Woodrow, I.E., Berry, J.A., 1987. A Model Predicting Stomatal Conductance and its Contribution to the Control of Photosynthesis under Different Environmental Conditions, in: Biggins, J. (Ed.), Progress in Photosynthesis Research. Springer Netherlands, Dordrecht, pp. 221–224. https://doi.org/10.1007/978-94-017-0519-6_48

Betts, A.K., 1986. A new convective adjustment scheme. Part I: Observational and theoretical basis. Quart J Royal Meteoro Soc 112, 677–691. https://doi.org/10.1002/qj.49711247307

Björnsson, H., Venegas, S.A., 1997. A manual of EOF and SVD analysis of climatic data, CCGCR Report No. 97-1. McGill University, Montréal, Québec.

Bonan, G.B., 2008. Soil moisture and the atmospheric boundary layer, in: Ecological Climatology: Concepts and Applications. Cambridge University Press, pp. 214–226.

Bonan, G.B., Oleson, K.W., Vertenstein, M., Levis, S., Zeng, X., Dai, Y., Dickinson, R.E., Yang, Z.-L., 2002. The Land Surface Climatology of the Community Land Model Coupled to the NCAR Community Climate Model*. J. Climate 15, 3123–3149. https://doi.org/10.1175/1520-0442(2002)015<3123:TLSCOT>2.0.CO;2

Brutsaert, W., 1982. Evaporation into the Atmosphere. Springer Netherlands, Dordrecht. https://doi.org/10.1007/978-94-017-1497-6

Cai, G., König, M., Carminati, A., Abdalla, M., Javaux, M., Wankmüller, F., Ahmed, M.A., 2024. Transpiration response to soil drying and vapor pressure deficit is soil texture specific. Plant Soil 500, 129–145. https://doi.org/10.1007/s11104-022-05818-2

Cammalleri, C., Vogt, J.V., Bisselink, B., De Roo, A., 2017. Comparing soil moisture anomalies from multiple independent sources over different regions across the globe. Hydrol. Earth Syst. Sci. 21, 6329–6343. https://doi.org/10.5194/hess-21-6329-2017

Chang, M., Liao, W., Wang, X., Zhang, Q., Chen, W., Wu, Z., Hu, Z., 2020. An optimal ensemble of the Noah-MP land surface model for simulating surface heat fluxes over a typical subtropical forest in South China. Agricultural and Forest Meteorology 281, 107815. https://doi.org/10.1016/j.agrformet.2019.107815

Chen, F., Dudhia, J., 2001. Coupling an Advanced Land Surface–Hydrology Model with the Penn State–NCAR MM5 Modeling System. Part I: Model Implementation and Sensitivity. Mon. Wea. Rev. 129, 569–585. https://doi.org/10.1175/1520-0493(2001)129<0569:CAALSH>2.0.CO;2

Chen, F., Janjić, Z., Mitchell, K., 1997. Impact of Atmospheric Surface-layer Parameterizations in the new Land-surface Scheme of the NCEP Mesoscale Eta Model. Boundary-Layer Meteorology 85, 391–421. https://doi.org/10.1023/A:1000531001463

Collins, W., Rasch, P., Boville, B., McCaa, J., Williamson, D., Kiehl, J., Briegleb, B., Bitz, C., Lin, S.-J., Zhang, M., Dai, Y., 2004. Description of the NCAR Community Atmosphere Model (CAM 3.0). UCAR/NCAR. https://doi.org/10.5065/D63N21CH

Dickinson, R.E., Shaikh, M., Bryant, R., Graumlich, L., 1998. Interactive Canopies for a Climate Model. J. Climate 11, 2823–2836. https://doi.org/10.1175/1520-0442(1998)011<2823:ICFACM>2.0.CO;2

Ek, M.B., Mitchell, K.E., Lin, Y., Rogers, E., Grunmann, P., Koren, V., Gayno, G., Tarpley, J.D., 2003. Implementation of Noah land surface model advances in the National Centers for Environmental Prediction operational mesoscale Eta model. J. Geophys. Res. 108, 2002JD003296. https://doi.org/10.1029/2002JD003296

Friedl, M.A., Sulla-Menashe, D., Tan, B., Schneider, A., Ramankutty, N., Sibley, A., Huang, X., 2010. MODIS Collection 5 global land cover: Algorithm refinements and characterization of new datasets. Remote Sensing of Environment 114, 168–182. https://doi.org/10.1016/j.rse.2009.08.016

Gan, Y., Liang, X., Duan, Q., Chen, F., Li, J., Zhang, Y., 2019. Assessment and Reduction of the Physical Parameterization Uncertainty for Noah-MP Land Surface Model. Water Resources Research 55, 5518–5538. https://doi.org/10.1029/2019WR024814

García-Valdecasas Ojeda, M., Gámiz-Fortis, S.R., Castro-Díez, Y., Esteban-Parra, M.J., 2017. Evaluation of WRF capability to detect dry and wet periods in Spain using drought indices. JGR Atmospheres 122, 1569–1594. https://doi.org/10.1002/2016JD025683

García-Valdecasas Ojeda, M., Gámiz-Fortis, S.R., Romero-Jiménez, E., Rosa-Cánovas, J.J., Yeste, P., Castro-Díez, Y., Esteban-Parra, M.J., 2021. Projected changes in the Iberian Peninsula drought characteristics. Science of The Total Environment 757, 143702. https://doi.org/10.1016/j.scitotenv.2020.143702

García-Valdecasas Ojeda, M., Rosa-Cánovas, J.J., Romero-Jiménez, E., Yeste, P., Gámiz-Fortis, S.R., Castro-Díez, Y., Esteban-Parra, M.J., 2020a. The role of the surface evapotranspiration in regional climate modelling: Evaluation and near-term future changes. Atmospheric Research 237, 104867. https://doi.org/10.1016/j.atmosres.2020.104867

García-Valdecasas Ojeda, M., Yeste, P., Gámiz-Fortis, S.R., Castro-Díez, Y., Esteban-Parra, M.J., 2020b. Future changes in land and atmospheric variables: An analysis of their couplings in the Iberian Peninsula. Science of The Total Environment 722, 137902. https://doi.org/10.1016/j.scitotenv.2020.137902

Gómez, I., Molina, S., Galiana-Merino, J.J., Estrela, M.J., Caselles, V., 2021. Impact of Noah-LSM Parameterizations on WRF Mesoscale Simulations: Case Study of Prevailing Summer Atmospheric Conditions over a Typical Semi-Arid Region in Eastern Spain. Sustainability 13, 11399. https://doi.org/10.3390/su132011399

Gómez-Navarro, J.J., Montávez, J.P., Jimenez-Guerrero, P., Jerez, S., García-Valero, J.A., González-Rouco, J.F., 2010. Warming patterns in regional climate change projections over the Iberian Peninsula. metz 19, 275–285. https://doi.org/10.1127/0941-2948/2010/0351

González-Zamora, Á., Sánchez, N., Pablos, M., Martínez-Fernández, J., 2019. CCI soil moisture assessment with SMOS soil moisture and in situ data under different environmental conditions and spatial scales in Spain. Remote Sensing of Environment 225, 469–482. https://doi.org/10.1016/j.rse.2018.02.010

Hersbach, H., Bell, B., Berrisford, P., Hirahara, S., Horányi, A., Muñoz-Sabater, J., Nicolas, J., Peubey, C., Radu, R., Schepers, D., Simmons, A., Soci, C., Abdalla, S., Abellan, X., Balsamo, G., Bechtold, P., Biavati, G., Bidlot, J., Bonavita, M., De Chiara, G., Dahlgren, P., Dee, D., Diamantakis, M., Dragani, R., Flemming, J., Forbes, R., Fuentes, M., Geer, A., Haimberger, L., Healy, S., Hogan, R.J., Hólm, E., Janisková, M., Keeley,

S., Laloyaux, P., Lopez, P., Lupu, C., Radnoti, G., De Rosnay, P., Rozum, I., Vamborg, F., Villaume, S., Thépaut, J., 2020. The ERA5 global reanalysis. Quart J Royal Meteoro Soc 146, 1999–2049. https://doi.org/10.1002/qj.3803

Hong, S.-Y., Dudhia, J., Chen, S.-H., 2004. A Revised Approach to Ice Microphysical Processes for the Bulk Parameterization of Clouds and Precipitation. Mon. Wea. Rev. 132, 103–120. https://doi.org/10.1175/1520-0493(2004)132<0103:ARATIM>2.0.CO;2

Hosseini, A., Mocko, D.M., Brunsell, N.A., Kumar, S.V., Mahanama, S., Arsenault, K., Roundy, J.K., 2022. Understanding the impact of vegetation dynamics on the water cycle in the Noah-MP model. Front. Water 4, 925852. https://doi.org/10.3389/frwa.2022.925852

Hu, W., Ma, W., Yang, Z., Ma, Y., Xie, Z., 2023a. Sensitivity Analysis of the Noah-MP Land Surface Model for Soil Hydrothermal Simulations Over the Tibetan Plateau. J Adv Model Earth Syst 15, e2022MS003136. https://doi.org/10.1029/2022MS003136

Hu, W., Ma, W., Yang, Z., Ma, Y., Xie, Z., 2023b. Sensitivity Analysis of the Noah-MP Land Surface Model for Soil Hydrothermal Simulations Over the Tibetan Plateau. J Adv Model Earth Syst 15, e2022MS003136. https://doi.org/10.1029/2022MS003136

Jacob, D., Petersen, J., Eggert, B., Alias, A., Christensen, O.B., Bouwer, L.M., Braun, A., Colette, A., Déqué, M., Georgievski, G., Georgopoulou, E., Gobiet, A., Menut, L., Nikulin, G., Haensler, A., Hempelmann, N., Jones, C., Keuler, K., Kovats, S., Kröner, N., Kotlarski, S., Kriegsmann, A., Martin, E., Van Meijgaard, E., Moseley, C., Pfeifer, S., Preuschmann, S., Radermacher, C., Radtke, K., Rechid, D., Rounsevell, M., Samuelsson, P., Somot, S., Soussana, J.-F., Teichmann, C., Valentini, R., Vautard, R., Weber, B., Yiou, P., 2014. EURO-CORDEX: new high-resolution climate change projections for European impact research. Reg Environ Change 14, 563–578. https://doi.org/10.1007/s10113-013-0499-2

Janjić, Z.I., 1994. The Step-Mountain Eta Coordinate Model: Further Developments of the Convection, Viscous Sublayer, and Turbulence Closure Schemes. Mon. Wea. Rev. 122, 927–945. https://doi.org/10.1175/1520-0493(1994)122<0927:TSMECM>2.0.CO;2

Jarvis, P.G., 1976. The interpretation of the variations in leaf water potential and stomatal conductance found in canopies in the field. Phil. Trans. R. Soc. Lond. B 273, 593–610. https://doi.org/10.1098/rstb.1976.0035

Jerez, S., López-Romero, J.M., Turco, M., Lorente-Plazas, R., Gómez-Navarro, J.J., Jiménez-Guerrero, P., Montávez, J.P., 2020. On the Spin-Up Period in WRF Simulations Over Europe: Trade-Offs Between Length and Seasonality. J Adv Model Earth Syst 12, e2019MS001945. https://doi.org/10.1029/2019MS001945

Jerez, S., Montavez, J.P., Gomez-Navarro, J.J., Jimenez-Guerrero, P., Jimenez, J., Gonzalez-Rouco, J.F., 2010. Temperature sensitivity to the land-surface model in MM5 climate simulations over the Iberian Peninsula. metz 19, 363–374. https://doi.org/10.1127/0941-2948/2010/0473

Kavvas, M.L., Delleur, J.W., 1975. Removal of periodicities by differencing and monthly mean subtraction. Journal of Hydrology 26, 335–353. https://doi.org/10.1016/0022-1694(75)90013-X

Khodayar, S., Sehlinger, A., Feldmann, H., Kottmeier, Ch., 2015. Sensitivity of soil moisture initialization for decadal predictions under different regional climatic conditions in Europe. Intl Journal of Climatology 35, 1899–1915. https://doi.org/10.1002/joc.4096

Klein, C., Bliefernicht, J., Heinzeller, D., Gessner, U., Klein, I., Kunstmann, H., 2017. Feedback of observed interannual vegetation change: a regional climate model analysis for the West African monsoon. Clim Dyn 48, 2837–2858. https://doi.org/10.1007/s00382-016-3237-x

Knist, S., Goergen, K., Buonomo, E., Christensen, O.B., Colette, A., Cardoso, R.M., Fealy, R., Fernández, J., García-Díez, M., Jacob, D., Kartsios, S., Katragkou, E., Keuler, K., Mayer, S., Van Meijgaard, E., Nikulin, G., Soares, P.M.M., Sobolowski, S., Szepszo, G., Teichmann, C., Vautard, R., Warrach-Sagi, K., Wulfmeyer, V., Simmer, C., 2017. Land-atmosphere coupling in EURO-CORDEX evaluation experiments. JGR Atmospheres 122, 79–103. https://doi.org/10.1002/2016JD025476

Koren, V., Schaake, J., Mitchell, K., Duan, Q. -Y., Chen, F., Baker, J.M., 1999. A parameterization of snowpack and frozen ground intended for NCEP weather and climate models. J. Geophys. Res. 104, 19569–19585. https://doi.org/10.1029/1999JD900232

Lawrence, D.M., Fisher, R.A., Koven, C.D., Oleson, K.W., Swenson, S.C., Bonan, G., Collier, N., Ghimire, B., Van Kampenhout, L., Kennedy, D., Kluzek, E., Lawrence, P.J., Li, F., Li, H., Lombardozzi, D., Riley, W.J., Sacks, W.J., Shi, M., Vertenstein, M., Wieder, W.R., Xu, C., Ali, A.A., Badger, A.M., Bisht, G., Van Den Broeke, M., Brunke, M.A., Burns, S.P., Buzan, J., Clark, M., Craig, A., Dahlin, K., Drewniak, B., Fisher, J.B., Flanner, M., Fox, A.M., Gentine, P., Hoffman, F., Keppel-Aleks, G., Knox, R., Kumar, S., Lenaerts, J., Leung, L.R., Lipscomb, W.H., Lu, Y., Pandey, A., Pelletier, J.D., Perket, J., Randerson, J.T., Ricciuto, D.M., Sanderson, B.M., Slater, A., Subin, Z.M., Tang, J., Thomas, R.Q., Val Martin, M., Zeng, X., 2019. The Community Land Model Version 5: Description of New Features, Benchmarking, and Impact of Forcing Uncertainty. J Adv Model Earth Syst 11, 4245–4287. https://doi.org/10.1029/2018MS001583

Li, J., Chen, F., Lu, X., Gong, W., Zhang, G., Gan, Y., 2020. Quantifying Contributions of Uncertainties in Physical Parameterization Schemes and Model Parameters to Overall Errors in Noah-MP Dynamic Vegetation Modeling. J Adv Model Earth Syst 12, e2019MS001914. https://doi.org/10.1029/2019MS001914

Li, J., Miao, C., Zhang, G., Fang, Y., Shangguan, W., Niu, G., 2022. Global Evaluation of the Noah-MP Land Surface Model and Suggestions for Selecting Parameterization Schemes. JGR Atmospheres 127, e2021JD035753. https://doi.org/10.1029/2021JD035753

Liang, X., Wood, E.F., Lettenmaier, D.P., 1996. Surface soil moisture parameterization of the VIC-2L model: Evaluation and modification. Global and Planetary Change 13, 195–206. https://doi.org/10.1016/0921-8181(95)00046-1

Liu, X., Chen, F., Barlage, M., Zhou, G., Niyogi, D., 2016. Noah-MP-Crop: Introducing dynamic crop growth in the Noah-MP land surface model. JGR Atmospheres 121. https://doi.org/10.1002/2016JD025597

Ma, N., 2023. Modeling land-atmosphere energy and water exchanges in the typical alpine grassland in Tibetan Plateau using Noah-MP. Journal of Hydrology: Regional Studies 50, 101596. https://doi.org/10.1016/j.ejrh.2023.101596

Miller, D.A., White, R.A., 1998. A conterminous United States multi-layer soil characteristics data set for regional climate and hydrology modeling, Earth Interactions 2.

Miralles, D.G., Gentine, P., Seneviratne, S.I., Teuling, A.J., 2019. Land–atmospheric feedbacks during droughts and heatwaves: state of the science and current challenges. Ann. N.Y. Acad. Sci. 1436, 19–35. https://doi.org/10.1111/nyas.13912

Miralles, D.G., Teuling, A.J., Van Heerwaarden, C.C., Vilà-Guerau De Arellano, J., 2014. Mega-heatwave temperatures due to combined soil desiccation and atmospheric heat accumulation. Nature Geosci 7, 345–349. https://doi.org/10.1038/ngeo2141

Neukam, D., Böttcher, U., Kage, H., 2016. Modelling Wheat Stomatal Resistance in Hourly Time Steps from Micrometeorological Variables and Soil Water Status. J Agronomy Crop Science 202, 174–191. https://doi.org/10.1111/jac.12133

Niu, G., Yang, Z., 2004. Effects of vegetation canopy processes on snow surface energy and mass balances. J. Geophys. Res. 109, 2004JD004884. https://doi.org/10.1029/2004JD004884

Niu, G., Yang, Z., Dickinson, R.E., Gulden, L.E., 2005. A simple TOPMODEL-based runoff parameterization (SIMTOP) for use in global climate models. J. Geophys. Res. 110, 2005JD006111. https://doi.org/10.1029/2005JD006111

Niu, G., Yang, Z., Dickinson, R.E., Gulden, L.E., Su, H., 2007. Development of a simple groundwater model for use in climate models and evaluation with Gravity Recovery and Climate Experiment data. J. Geophys. Res. 112, 2006JD007522. https://doi.org/10.1029/2006JD007522

Niu, G.-Y., Yang, Z.-L., 2006. Effects of Frozen Soil on Snowmelt Runoff and Soil Water Storage at a Continental Scale. Journal of Hydrometeorology 7, 937–952. https://doi.org/10.1175/JHM538.1

Niu, G.-Y., Yang, Z.-L., Mitchell, K.E., Chen, F., Ek, M.B., Barlage, M., Kumar, A., Manning, K., Niyogi, D., Rosero, E., Tewari, M., Xia, Y., 2011. The community Noah land surface model with multiparameterization options (Noah-MP): 1. Model description and evaluation with local-scale measurements. J. Geophys. Res. 116, D12109. https://doi.org/10.1029/2010JD015139

Oleson, K., Lawrence, D., Bonan, G., Flanner, M., Kluzek, E., Lawrence, P., Levis, S., Swenson, S., Thornton, P., Dai, A., Decker, M., Dickinson, R., Feddema, J., Heald, C., Hoffman, F., Lamarque, J.-F., Mahowald, N., Niu, G.-Y., Qian, T., Randerson, J., Running, S., Sakaguchi, K., Slater, A., Stockli, R., Wang, A., Yang, Z.-L., Zeng, Xiaodong, Zeng, Xubin, 2010. Technical Description of version 4.0 of the Community Land Model (CLM). UCAR/NCAR. https://doi.org/10.5065/D6FB50WZ

Pampuch, L.A., Negri, R.G., Loikith, P.C., Bortolozo, C.A., 2023. A Review on Clustering Methods for Climatology Analysis and Its Application over South America. IJG 14, 877–894. https://doi.org/10.4236/ijg.2023.149047

Peral García, C., Navascués Fernández-Victorio, B., Ramos Calzado, P., 2017. Serie de precipitación diaria en rejilla con fines climáticos. Agencia Estatal de Meteorología. https://doi.org/10.31978/014-17-009-5

Perez-Priego, O., El-Madany, T.S., Migliavacca, M., Kowalski, A.S., Jung, M., Carrara, A., Kolle, O., Martín, M.P., Pacheco-Labrador, J., Moreno, G., Reichstein, M., 2017. Evaluation of eddy covariance latent heat fluxes with independent lysimeter and sapflow estimates in a Mediterranean savannah ecosystem. Agricultural and Forest Meteorology 236, 87–99. https://doi.org/10.1016/j.agrformet.2017.01.009

Pleim, J.E., 2007. A Combined Local and Nonlocal Closure Model for the Atmospheric Boundary Layer. Part I: Model Description and Testing. Journal of Applied Meteorology and Climatology 46, 1383–1395. https://doi.org/10.1175/JAM2539.1

Pool, S., Vis, M., Seibert, J., 2018. Evaluating model performance: towards a non-parametric variant of the Kling-Gupta efficiency. Hydrological Sciences Journal 63, 1941–1953. https://doi.org/10.1080/02626667.2018.1552002

Preisendorfer, R.W., 1988. Principal component analysis in meteorology and oceanography. Elsevier Sci 425.

Qi, Y., Zhang, Q., Hu, S., Wang, R., Wang, H., Zhang, K., Zhao, H., Zhao, F., Chen, F., Yang, Y., Tang, G., Hu, Y., 2023. Applicability of stomatal conductance models comparison for persistent water stress processes of spring maize in water resources limited environmental zone. Agricultural Water Management 277, 108090. https://doi.org/10.1016/j.agwat.2022.108090

Ridal, M., Bazile, E., Le Moigne, P., Randriamampianina, R., Schimanke, S., Andrae, U., Berggren, L., Brousseau, P., Dahlgren, P., Edvinsson, L., El-Said, A., Glinton, M., Hagelin, S., Hopsch, S., Isaksson, L., Medeiros, P., Olsson, E., Unden, P., Wang, Z.Q., 2024. CERRA , the Copernicus European Regional Reanalysis system. Quart J Royal Meteoro Soc 150, 3385–3411. https://doi.org/10.1002/qj.4764

Rousseeuw, P.J., 1987. Silhouettes: A graphical aid to the interpretation and validation of cluster analysis. Journal of Computational and Applied Mathematics 20, 53–65. https://doi.org/10.1016/0377-0427(87)90125-7

Sakaguchi, K., Zeng, X., 2009. Effects of soil wetness, plant litter, and under-canopy atmospheric stability on ground evaporation in the Community Land Model (CLM3.5). J. Geophys. Res. 114, 2008JD010834. https://doi.org/10.1029/2008JD010834

Schaake, J.C., Koren, V.I., Duan, Q., Mitchell, K., Chen, F., 1996. Simple water balance model for estimating runoff at different spatial and temporal scales. J. Geophys. Res. 101, 7461–7475. https://doi.org/10.1029/95JD02892

Schimanke, S., Ridal, M., Le Moigne, P., Berggren, L., Undén, P., Randriamampianina, R., Andrea, U., Bazile, E., Bertelsen, A., Brousseau, P., Dahlgren, P., Edvinsson, L., El Said, A., Glinton, M., Hopsch, S., Isaksson, L., Mladek, R., Olsson, E., Verrelle, A., Wang, Z.Q., 2021. CERRA sub-daily regional reanalysis data for Europe on single levels from 1984 to present. https://doi.org/10.24381/CDS.622A565A

Sellers, P.J., Heiser, M.D., Hall, F.G., 1992. Relations between surface conductance and spectral vegetation indices at intermediate (100 m$^2$ to 15 km$^2$ ) length scales. J. Geophys. Res. 97, 19033–19059. https://doi.org/10.1029/92JD01096

Seneviratne, S.I., Corti, T., Davin, E.L., Hirschi, M., Jaeger, E.B., Lehner, I., Orlowsky, B., Teuling, A.J., 2010. Investigating soil moisture–climate interactions in a changing climate: A review. Earth-Science Reviews 99, 125–161. https://doi.org/10.1016/j.earscirev.2010.02.004

Seneviratne, S.I., Lüthi, D., Litschi, M., Schär, C., 2006. Land–atmosphere coupling and climate change in Europe. Nature 443, 205–209. https://doi.org/10.1038/nature05095

Serrano-Ortiz, P., Kowalski, A.S., Domingo, F., Rey, A., Pegoraro, E., Villagarcía, L., Alados-Arboledas, L., 2007. Variations in daytime net carbon and water exchange in a montane shrubland ecosystem in southeast Spain. Photosynt. 45, 30–35. https://doi.org/10.1007/s11099-007-0005-5

Skamarock, W.C., Klemp, J.B., Jimy Dudhia, Gill, D.O., Barker, D.M., Duda, M.G., Huang, X.-Y., Wang, W., Powers, J.G., 2008. A Description of the Advanced Research WRF Version 3. https://doi.org/10.13140/RG.2.1.2310.6645

Solano-Farias, F., García-Valdecasas Ojeda, M., Donaire-Montaño, D., Rosa-Cánovas, J.J., Castro-Díez, Y., Esteban-Parra, M.J., Gámiz-Fortis, S.R., 2024. Assessment of physical schemes for WRF model in convection-permitting mode over southern Iberian Peninsula. Atmospheric Research 299, 107175. https://doi.org/10.1016/j.atmosres.2023.107175

Torres-Rojas, L., Vergopolan, N., Herman, J.D., Chaney, N.W., 2022. Towards an Optimal Representation of Sub-Grid Heterogeneity in Land Surface Models. Water Resources Research 58, e2022WR032233. https://doi.org/10.1029/2022WR032233

Verseghy, D.L., 1991. Class—A Canadian land surface scheme for GCMS. I. Soil model. Intl Journal of Climatology 11, 111–133. https://doi.org/10.1002/joc.3370110202

von Storch, H., Navarra, A., 1995. Analysis of Climate Variability: Applications of Statistical Techniques. Springer-Verlag.

Wallace, J.M., Hobbs, P.V., 2006. Atmospheric Science. Elsevier. https://doi.org/10.1016/C2009-0-00034-8

Xue, Y., Sellers, P.J., Kinter, J.L., Shukla, J., 1991. A Simplified Biosphere Model for Global Climate Studies. J. Climate 4, 345–364. https://doi.org/10.1175/1520-0442(1991)004<0345:ASBMFG>2.0.CO;2

Yang, Q., Dan, L., Lv, M., Wu, J., Li, W., Dong, W., 2021. Quantitative assessment of the parameterization sensitivity of the Noah-MP land surface model with dynamic vegetation using ChinaFLUX data. Agricultural and Forest Meteorology 307, 108542. https://doi.org/10.1016/j.agrformet.2021.108542

Yang, Z.-L., Dickinson, R.E., 1996. Description of the Biosphere-Atmosphere Transfer Scheme (BATS) for the Soil Moisture Workshop and evaluation of its performance. Global and Planetary Change 13, 117–134. https://doi.org/10.1016/0921-8181(95)00041-0

Yang, Z.-L., Dickinson, R.E., Robock, A., Vinnikov, K.Y., 1997. Validation of the Snow Submodel of the Biosphere–Atmosphere Transfer Scheme with Russian Snow Cover and Meteorological Observational Data. J. Climate 10, 353–373. https://doi.org/10.1175/1520-0442(1997)010<0353:VOTSSO>2.0.CO;2

Yang, Z.-L., Niu, G.-Y., Mitchell, K.E., Chen, F., Ek, M.B., Barlage, M., Longuevergne, L., Manning, K., Niyogi, D., Tewari, M., Xia, Y., 2011. The community Noah land surface model with multiparameterization options (Noah-MP): 2. Evaluation over global river basins. J. Geophys. Res. 116, D12110. https://doi.org/10.1029/2010JD015140

Yeste, P., García-Valdecasas Ojeda, M., Gámiz-Fortis, S.R., Castro-Díez, Y., Esteban-Parra, M.J., 2020. Integrated sensitivity analysis of a macroscale hydrologic model in the north of the Iberian Peninsula. Journal of Hydrology 590, 125230. https://doi.org/10.1016/j.jhydrol.2020.125230

You, Y., Huang, C., Yang, Z., Zhang, Y., Bai, Y., Gu, J., 2020. Assessing Noah-MP Parameterization Sensitivity and Uncertainty Interval Across Snow Climates. JGR Atmospheres 125, e2019JD030417. https://doi.org/10.1029/2019JD030417

Zhang, G., Chen, F., Chen, Y., Li, J., Peng, X., 2020. Evaluation of Noah-MP Land-Model Uncertainties over Sparsely Vegetated Sites on the Tibet Plateau. Atmosphere 11, 458. https://doi.org/10.3390/atmos11050458

Zhang, G., Chen, F., Gan, Y., 2016. Assessing uncertainties in the Noah-MP ensemble simulations of a cropland site during the Tibet Joint International Cooperation program field campaign. JGR Atmospheres 121, 9576–9596. https://doi.org/10.1002/2016JD024928

Zhang, G., Li, J., Zhou, G., Cai, X., Gao, W., Peng, X., Chen, Y., 2021. Effects of Mosaic Representation of Land Use/Land Cover on Skin Temperature and Energy Fluxes in Noah-MP Land Surface Model Over China. JGR Atmospheres 126, e2021JD034542. https://doi.org/10.1029/2021JD034542

Zhang, X., Chen, L., Ma, Z., Duan, J., Dai, D., Zhang, H., 2022. Effects of the surface coupling strength in the WRF/Noah-MP model on regional climate simulations over China. Clim Dyn 59, 331–355. https://doi.org/10.1007/s00382-021-06129-5

Zheng, H., Yang, Z., Lin, P., Wei, J., Wu, W., Li, L., Zhao, L., Wang, S., 2019. On the Sensitivity of the Precipitation Partitioning Into Evapotranspiration and Runoff in Land Surface Parameterizations. Water Resources Research 55, 95–111. https://doi.org/10.1029/2017WR022236

**FIGURE CAPTIONS**

**Fig 1** (a) Domain configuration for WRF simulations: the parent domain (d01) corresponding to the EURO-CORDEX region at 0.44º (approximately 50 km) spatial resolution and the child domain (d02) centered on the Iberian Peninsula (IP) at 0.08º spatial resolution (approximately 10 km), (b) altitude in the study region in relation to the location of the three FLUXNET stations used to validate the experiments, (c), land cover types across the IP according to the modified IGBP MODIS 20-category vegetation classification (ENF: evergreen needleleaf forest; DBF: deciduous broadleaf forest; MF: mixed forest; OS: open shrubland; WS: woody savanna; S: savanna; G: grassland; PW: permanent wetland; C: cropland; UBU: urban and built-up; C/NVM: cropland/natural vegetation mosaic; BSV: barely/sparsely vegetated; W: water) and (d) dominant soil textures appearing in the IP from the hybrid STATSGO/FAO 16-classes categories (Sa: sand; SLo: sandy loam; Lo: loam, SaCLo:sandy clay loam, CLo: clay loam, and C:clay).

**Fig 2** Experiments carried out with the different Noah-MP configurations. The 35 parameter combinations are represented in rows and each of the schemes in columns. The last column also shows the cluster to which the experiment belongs. For the different options, the asterisks indicate default options.

**Fig 3** Spatial mean of the annual precipitation anomalies for the IP expressed in mm year$^{-1}$. Climatology from the ROCIO-IBEB dataset (Peral García et al., 2017), using the period 1961-2022.

**Fig 4** Mean bias of the different experiments (from s0 to s34) in soil moisture content (SMOIS) for spring (MAM, March-April-May) 2010 (i.e., the wet year) compared to CERRA-Land. The background color of each map indicates the cluster to which the experiment belongs.

**Fig 5** The same as Fig 4 but for (a) the sensible heat flux (SH) and the latent heat flux (LH).

**Fig 6** Temporal correlation between (a) SH and (b) LH of each experiment and CERRA-Land during 2005 (i.e., the dry year). The background color of each map indicates the cluster to which the experiment belongs.

**Fig 7** Annual cycle of monthly mean SH (W m$^{-2}$), LH (W m$^{-2}$), and SMOIS standardized bias (dimensionless) at the three FLUXNET stations used in this evaluation for (a) 2005 and (b) 2010. Results for all experiments (dashed colored lines), for the cluster means (solid colored lines),

CERRA-Land (black dashed line) and stations (solid black line) are shown for each point and variable.

**Fig 8** Temporal correlation between SH and LH as a measure of ground-atmosphere coupling for each of the experiments and CERRA-Land in (a) 2005 and (b) 2010. The spatial pattern (r) as well as the root mean square error (RMSE) is shown in the lower right corner of each plot within the panel. The background color of each map indicates the cluster to which the experiment belongs.

**Fig 9** Clusters with the best KGE rank for each grid point in 2005 and 2010 for a) SH and b) LH.

**Table 1** WRF/Noah-MP parameterization options considered in this study. For each parameterization a brief description is given of each option.

| Parameterizations and description | Options |
|---|---|
| **Dynamic Vegetation (DVEG):** it determines how leaf area index (LAI) and greenness vegetation fraction (GVF) are calculated. | OFF (1, OFF-GVF): seasonally-varying LAI specified through a look-up table, and GVF corresponds to the shadow fraction (SHDFAC) (Niu et al., 2011, Yang et al., 2011). <br> ON (2, ON-Dickison): LAI and GVF are predicted using the Dickinson et al. (1998) model. <br> OFF (3, OFF-Calc): seasonally-varying LAI specified through a look-up table and GVF is calculated. <br> OFF (4, OFF-MVF) [default option]: seasonally-varying LAI specified through a look-up table. GVF is the maximum vegetation fraction (MVF) of the dominant vegetation type in each cell. <br> ON (5, ON-MVF): LAI is predicted through the Dickinson et al. (1998) model, and GVF is the MVF of the dominant vegetation for each grid cell. |
| **Canopy Stomatal Resistance (CRS):** it adjusts stomatal resistance per LAI unit (i.e., canopy resistance) and is related to the photosynthesis rate. | Ball-Berry (1, Ball-Berry) [default option]: empirical model linking photosynthesis and transpiration by relating stomatal conductance to $CO_2$ exchange rate through stomata. Useful for analyzing gas and carbon exchanges and the vegetation response to climate change (Ball et al., 1987). <br> Jarvis (2, Jarvis): semi-empirical formulation based on environmental factors (soil moisture, atmospheric temperature, radiation availability, and vapor pressure deficit) (Jarvis, 1976). |
| **Surface exchange coefficient for heat (SFC):** it modulates energy, water, and momentum exchanges processes. Therefore, it affects canopy evaporation, soil evaporation, transpiration, and snow surface energy balance as well as the interchange of latent and sensible heat fluxes. | Monin-Obukhov (1, M-O) [default option]: it uses the Monin-Obukhov length as a measure of atmospheric stability. Useful for simulations with more details in surface roughness or atmospheric stability (Brutsaert, 1982). <br> Original Noah (2, Chen97): empirical formulation applied to represent the mean planetary layer roughness and global simulations where a general representation of the land-atmosphere processes is enough, reducing thus the computational cost (Chen et al., 1997). |
| **Soil moisture factor controlling stomatal resistance (BTR):** it determines the effect of soil moisture on stomatal resistance through the soil moisture factor ($\beta$) which modulates the stomatal resistance, and therefore, transpiration. | Noah type (1, Noah) [default option]: the soil moisture factor ($\beta$) controlling stomatal resistance is calculated as a function of soil moisture calculated using a simplified Noah LSM model. Useful when we are interested in representing average conditions without complex details (Chen and Dudhia, 2001). <br> CLM type (2, CLM): it is based on the community land model (CAM) and uses photosynthetic processes. More complex than Noah Type with $\beta$ using matric potential. (Oleson et al., 2010). <br> SSiB type (3, SSiB): based on the simple biosphere model. It relates stomatal resistance with a $\beta$ based on matric potential Useful in semi-arid and desertification conditions (Xue et al., 1991). |
| **Runoff and groundwater (RUN):** it modulates the water movement (surface and subsurface) through the soil. Critical for the hydrological cycle. | Topography-based hydrological model (TOPMODEL) with groundwater (1, GW): Full topography-based groundwater and water table dynamics. Despite being computationally demanding with detailed data requirements, it is more realistic (Niu et al., 2007). <br> TOPMODEL with equilibrium water table (2, EQWT): simplified water table model. Useful for global scale in where subsurface runoff is a result of an exponential function of water table depth and a single coefficient, making the model more feasible to apply coupled to GCMs (Niu et al., 2005). <br> Original surface and subsurface runoff (3, SR) [default option]: infiltration-excess-based surface runoff scheme with a gravitational free-drainage subsurface runoff scheme (Schaake et al., 1996). <br> Biosphere-atmosphere transfer scheme (BATS) runoff scheme (4, BATS): surface runoff defined as a $4^{th}$ power function of the top 2 m wetness and subsurface runoff as a gravitational free drainage (Yang and Dickinson, 1996). |
| **Supercooled liquid water (FRZ):** it controls how the model simulates the water behavior when it is in a supercooled liquid state. | No iteration (1, NoIt) [default option]: There is no iterative correction to consider variations in the state of supercooled liquid water, (Niu and Yang, 2006). <br><br> Koren's iteration (2, Koren): The model iterates over the thermodynamic calculations an extra term that accounts for the increased interface between soil particles and liquid water (Koren et al., 1999). |
| **Frozen soil permeability (INF):** it regulates how water infiltrates (if more quickly or slowly) through the soil (i.e., soil hydraulic properties). | Linear effects, more permeable (1, Linear) [default option]: it is modeled as a linear function of soil moisture (Niu and Yang, 2006). <br><br> Non-linear effects, less permeable (2, Non-Linear): soil permeability is parametrized with a more complex relationship using the liquid water volume (Koren et al., 1999). |
| **Radiative transfer (RAD):** it adjusts how canopy radiation transfer is treated. Based on two-stream models and the difference between options is related to the way in which they treat the gaps that occurred tree crowns. | Modified two-stream scheme (1, ModTS): simplified version of radiative transfer model in which the interaction is bidirectional (Niu and Yang, 2004). <br><br> Two-stream scheme applied to grid-cell (2, GridTS): standard two stream to the grid without vegetation (Niu and Yang, 2004). <br><br> Two-stream scheme applied to vegetated fraction (3, VegTS) [default option]: it separates the grid into vegetation and no-vegetation fractions and apply it two-stream only to vegetation areas (Niu and Yang, 2004). |
| **Snow surface albedo (ALB):** it modulates the surface snow albedo. | BATS (1, BATS): it calculates snow surface albedo for both direct and diffuse radiation over the visible and near-infrared bands, taking into consideration fresh snow albedo, fluctuations in snow age, solar zenith angle, grain size growth, and contaminants (Yang et al., 1997). <br><br> Canadian Land Surface Scheme (2, CLASS) [default option]: snow surface albedo is computed using the fresh snow albedo and snow age (Verseghy, 1991). |
| **Lower boundary condition for soil temperature (TBOT):** It determines the treatment of temperature at the soil column's lower boundary. | Zero heat flux (1, ZeroHF): it assumes that there is no heat transfer over the soil column's bottom border, setting a constant temperature value at depth without heat transfer across the soil. It is simple and computationally efficient (Niu et al., 2011). <br><br> TBOT at 8 m from input file (2, TBOT8m) [default option]: it allows a fixed soil temperature at 8 meters depth, which is provided as an input (Ek et al., 2003). |
| **Surface resistant to evaporation (RSF):** It controls the ground resistance to evaporation/sublimation, and directly influences the amount of water vapor that can escape into the atmosphere (i.e., soil evaporation). | Sakaguchi and Zeng method (1, SZ19) [default option]: surface resistance based on plant litter cover, water vapor transfer, and under-canopy atmospheric stability. Accurate for wet soil with dense vegetation (Sakaguchi and Zeng, 2009). <br><br> Sellers's method (2, Sellers): Empirical method to adjust the RSF based on the percentage of snow-covered ground and the topsoil layer's soil moisture content (Sellers et al., 1992). <br><br> Adjusted Sellers's to decrease RSF for wet soil (3, AS-Wet): as Sellers's but with empirical adjust for wet soils. Thus, uncertainties in wet soil are corrected (Sellers et al., 1992). |

**Table 2** Clusters and experiments that compose them as along with the options of the parameterizations that they have in common. Asterisks denote that the cluster uses the default option of the parameterization.

| Clusters | Experiments | Common configuration | Differences within the cluster |
|---|---|---|---|
| A | s0A, s6A, s21A, and s30A | CRS = Ball-Berry*<br>SFC = M-O*<br>FRZ = NoIt*<br>RAD = VegTS*<br>RSF = SZ19* | - DVEG: s0A and s30A calculate LAI while s6A and s21A obtain LAI through a look-up table.<br>- BTR: s0A and s6A use Noah vs. s21A and s30A use CLM.<br>- RUN: s0A and s30A use SR option while s6A and s21A use GW.<br>- INF: all experiments use Linear except the experiment s30A which uses NonLinear.<br>- ALB: all experiments use CLASS except s6A which uses BATS.<br>- TBOT: s0A and s30A use TBOT8m while s6A and s21A use ZeroHF. |
| B | s1B, s5B, s7B, s8B, s9B, and s11B | DVEG = On<br>CRS = Ball-Berry*<br>SFC = M-O*<br>BTR = Noah*<br>INF = Linear*<br>ALB = BATS<br>TBOT = ZeroHF<br>RSF = SZ19* | - DVEG: LAI is calculated in all experiments. However, s11B uses the dominant vegetation to obtain GVF and other experiments calculate it.<br>- RUN: all experiments use GW except s8B which uses BATS.<br>- FRZ: all experiments use NoIt except s9B which uses Koren.<br>- RAD: all experiments use ModTS except s5B which uses GridTS. |
| C | s2C and s22C | DVEG = Off<br>CRS = Ball-Berry*<br>SFC = M-O*<br>BTR = Noah*<br>RUN = GW<br>FRZ = NoIt*<br>INF = Linear*<br>RAD = ModTS<br>ALB = BATS<br>TBOT = ZeroHF<br>RSF = SZ19* | - DVEG: while s2C calculates GVF, s22C GVF corresponds to the shadow fraction. |
| D | s3D, s10D, s13D, s17D, s18D, and s25D | DVEG = On<br>CRS = Ball-Berry*<br>SFC = M-O*<br>FRZ = NoIt*<br>RAD = ModTS<br>RSF = SZ19* | - DVEG: s3D, s10D, s17D, s18D, and s25D calculate GVF, but s13D uses the maximum vegetation fraction.<br>- BTR: s3D and s13D use CLM while s10D, s13D, s17D, s18D, and s25D use SSiB.<br>- RUN: all experiments use GW except s25D which uses SR.<br>- INF: all experiments use Linear except s11D which uses NonLinear.<br>- ALB: experiments s3D, s10D, s17D use BATS and s18D and s25D CLASS.<br>- TBOT: all experiments use ZeroHF except s17D which uses TBOT8m. |
| E | s4E, s16E, and s32E | DVEG = On<br>CRS = Ball-Berry*<br>SFC = Chen97<br>BTR = Noah*<br>RUN = GW<br>FRZ = NoIt*<br>INF = Linear*<br>RSF = SZ19* | - ALB: s4E uses BATS while s16E and s32E use CLASS.<br>- RAD: s4E uses ModTS, s16E GridTS and s32E VegTS.<br>- TBOT: s4E and s32E use ZeroHF, while s16E uses TBOT8m. |
| F | s12F, s14F, and s15F | DVEG = On-MVF<br>CRS = Ball-Berry*<br>SFC = M-O* | - BTR: s12F uses Noah while s14F and s15F use CLM.<br>- RUN: GW (s12F and s14F) or BATS (s15F). |

| | | RAD = ModTS<br>ALB = BATS<br>TBOT = ZeroHF<br>RSF = AS-Wet | - FRZ: s12F and s14F use NoIt*, while s15F uses Koren.<br>- INF: s12F uses Linear* and s14F and s15F use NonLinear. |
|---|---|---|---|
| G | s19G, s20G, s23G, s24G, and s29G | DVEG = ON-Dickinson<br>CRS = Ball-Berry*<br>SFC = M-O*<br>RAD = VegTS*<br>ALB = CLASS* | - BTR: all experiments use SSiB except s29G which uses CLM.<br>- RUN: experiments s19G, s20G, and s29G use GW, while s23G and s24G use SR.<br>- FRZ: all experiments have the NoIt option except s29G which uses Koren.<br>- INF: all experiments use Linear except s29G which uses NonLinear.<br>- TBOT: experiments s19G and s24G use ZeroHF while s20G, s23G and s29G use TBOT8m.<br>- RSF: All experiments use SZ19 except s29G which uses AS-Wet. |
| H | s26H | DVEG = On-Dickinson<br>CRS = Jarvis<br>SFC = Chen97<br>BTR = Noah*<br>RUN = EQWT<br>FRZ = NoIt*<br>INF = Linear*<br>RAD = VegTS*<br>ALB = CLASS*<br>TBOT = ZeroHF<br>RSF = SZ19* | |
| I | s27I, s28I, and s31I | DVEG = Off<br>CRS = Jarvis<br>SFC = Chen97<br>RAD = VegTS*<br>ALB = CLASS*<br>FRZ = Koren | - DVEG: all experiments have OFF DVEG options, but while s27I and s28I calculate GVF, s31I GVF is the maximum vegetation fraction,<br>- BTR: s27I and s31I use Noah while s28I uses CLM.<br>- RUN: GW (s31I) or BATS (s27I and s28I).<br>- INF: s27I and s28I use NonLinear while s31I uses Linear.<br>- TBOT: s27I and s28I use the TBOT TBOT8m while s31I uses ZeroHF.<br>- RSF: s27I and s28I use AS-Wet while s31I uses Sellers. |
| J | s33J and s34J | DVEG = On<br>CRS = Ball-Berry*<br>SFC = M-O*<br>RUN = EQWT<br>FRZ = NoIt* | - DVEG: all experiments have ON DVEG options, but while s33J calculate GVF, s34J uses the maximum vegetation fraction.<br>- BTR: s33J uses CLM and s34J SSiB.<br>- INF: s33J uses NonLinear INF and s34J Linear.<br>- RAD: s33J uses ModTS and s34J VegTS.<br>- ALB: Experiment s33J uses BATS and s34J CLASS.<br>- RSF: s33J uses AS-Wet and s34J SZ19. |

**Table 3** Percentage of area with the best results for each cluster as well as the average amplitude for the SH and LH variables separately

| Cluster | Area (%) 2005 | | Area (%) 2010 | | Amplitude (%) 2005 | | Amplitude (%) 2010 | |
|---|---|---|---|---|---|---|---|---|
| | SH | LH | SH | LH | SH | LH | SH | LH |
| A | 13.41 | 2.60 | 14.28 | 3.05 | 41.44 | 72.55 | 42.96 | 66.85 |
| B | 8.87 | 3.10 | 8.53 | 3.52 | 30.30 | 33.49 | 31.54 | 38.14 |
| C | 5.32 | 3.54 | 5.06 | 6.31 | 13.57 | 16.68 | 11.12 | 12.34 |
| D | 3.64 | 5.32 | 2.54 | 2.84 | 55.23 | 46.15 | 60.21 | 61.76 |
| E | 10.78 | 18.28 | 6.26 | 21.72 | 30.75 | 16.19 | 34.48 | 14.84 |
| F | 20.12 | 24.01 | 13.76 | 19.45 | 20.09 | 43.76 | 23.32 | 33.31 |
| G | 14.70 | 7.56 | 17.28 | 5.67 | 31.28 | 42.90 | 38.72 | 47.98 |
| H | 0.85 | 0.02 | 1.03 | 0.02 | 0 | 0 | 0 | 0 |
| I | 20.44 | 35.40 | 30.60 | 37.30 | 22.34 | 29.35 | 18.16 | 24.70 |
| J | 1.87 | 0.17 | 0.64 | 0.12 | 7.40 | 20.29 | 4.07 | 55.92 |

**David Donaire-Montaño**: Methodology, Software, Validation, Formal Analysis, Writing – Original Draft.

**Matilde García-Valdecasas Ojeda**: Conceptualization, Software, Supervision, Writing – Original Draft, Writing – Review & Editing, Visualization, Investigation.

**Nicolás Tacoronte**: Methodology, Software.

**Juanjo José Rosa-Cánovas**: Software, Investigation.

**Yolanda Castro-Díez**: Investigation, Conceptualization, Writing – Review & Editing, Supervision.

**María Jesús Esteban-Parra**: Investigation, Conceptualization, Writing – Review & Editing, Supervision, Funding acquisition.

**Sonia Raquel Gámiz-Fortis**: Resources, Investigation, Writing – Review & Editing, Supervision, Funding acquisition, Project administration.

**Declaration of interests**

☒ The authors declare that they have no known competing financial interests or personal relationships that could have appeared to influence the work reported in this paper.

☐ The authors declare the following financial interests/personal relationships which may be considered as potential competing interests:

# Highlights

5. A clustering method is proposed to identify differences in Noah-MP parameterizations.

6. Noah-MP parameterizations impact heat flux performance and land-atmosphere interactions.

7. RUN, SFC, BTR, CRS, and RSF are key parameterizations to simulate heat fluxes in the IP.

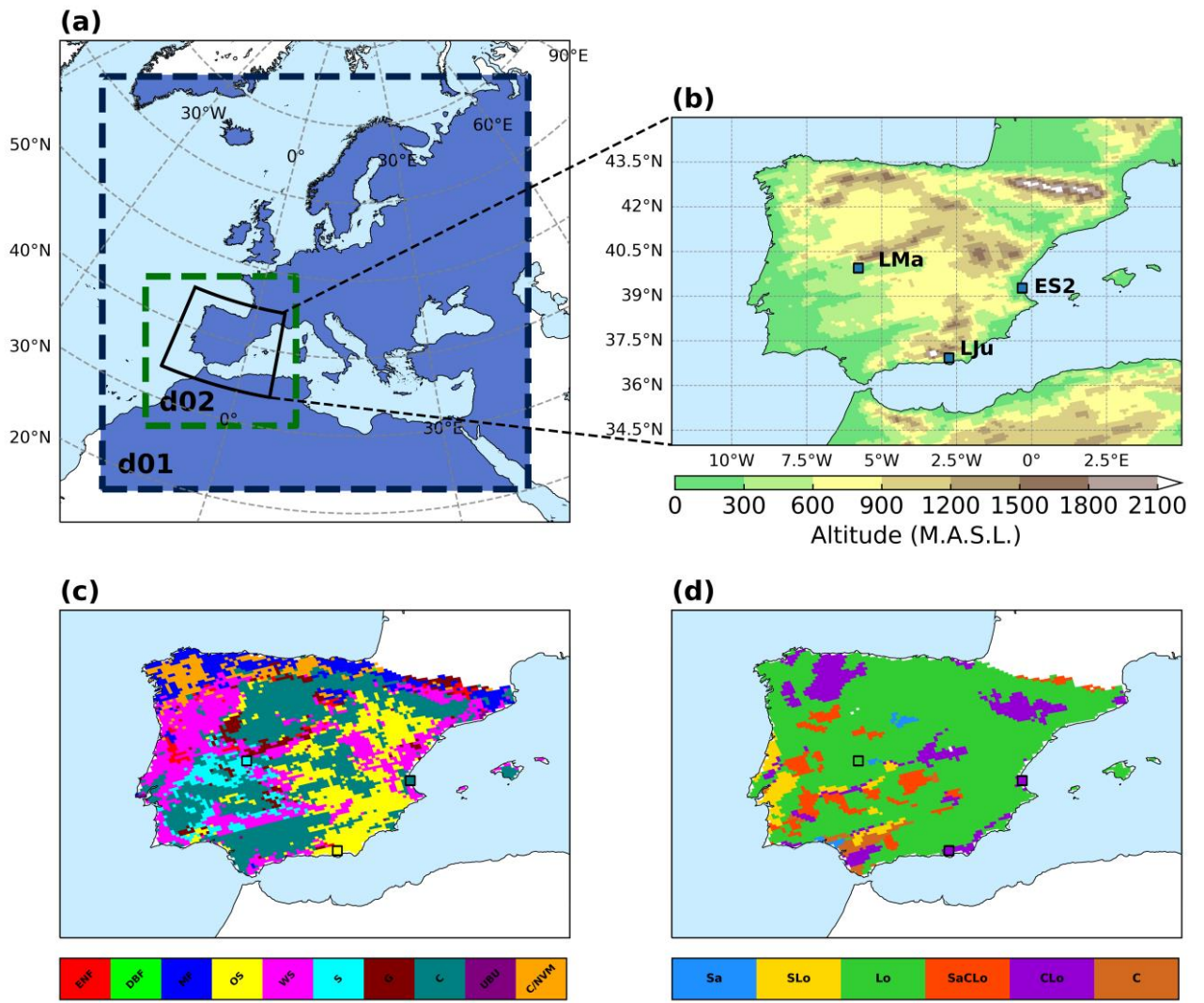

**Figure 1**

Figure 2

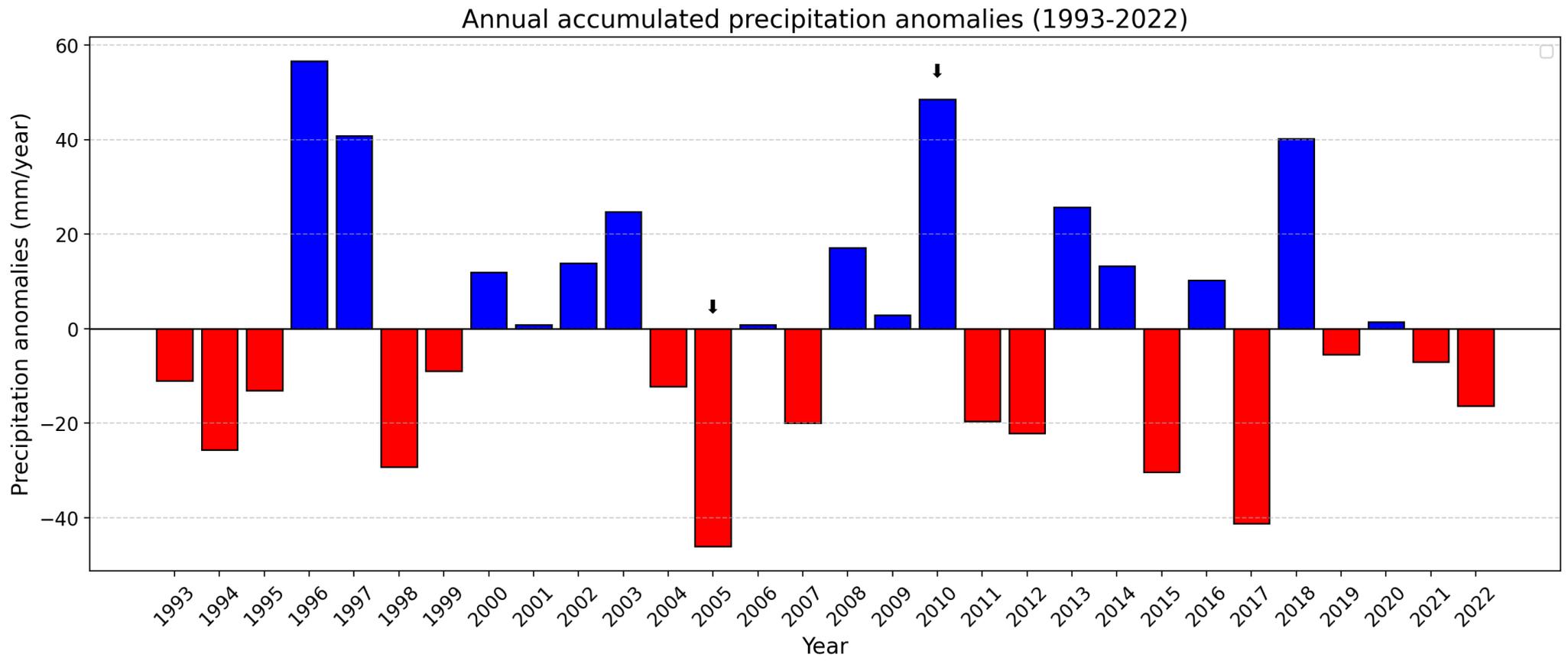

**Figure 3**

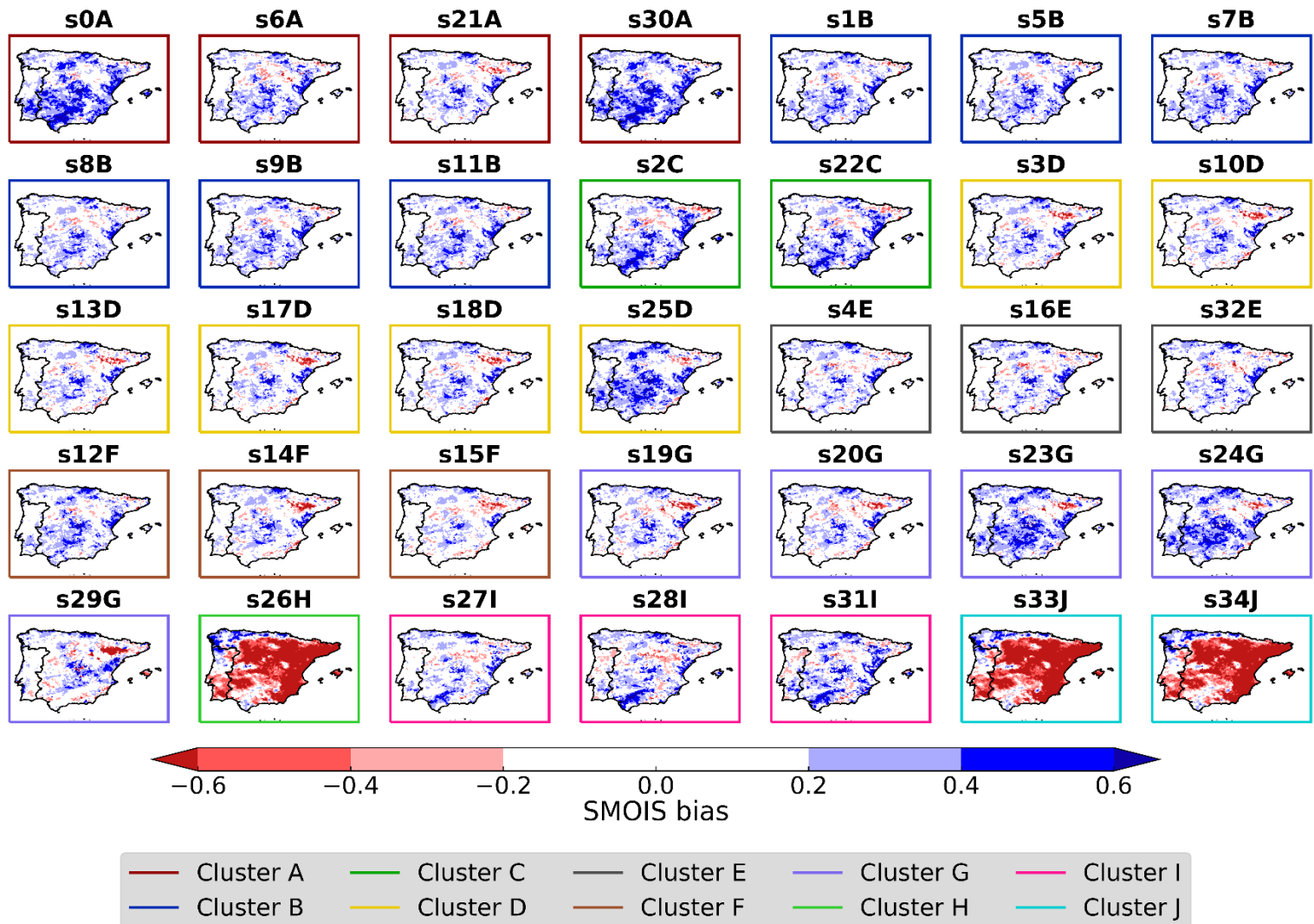

**Figure 4**

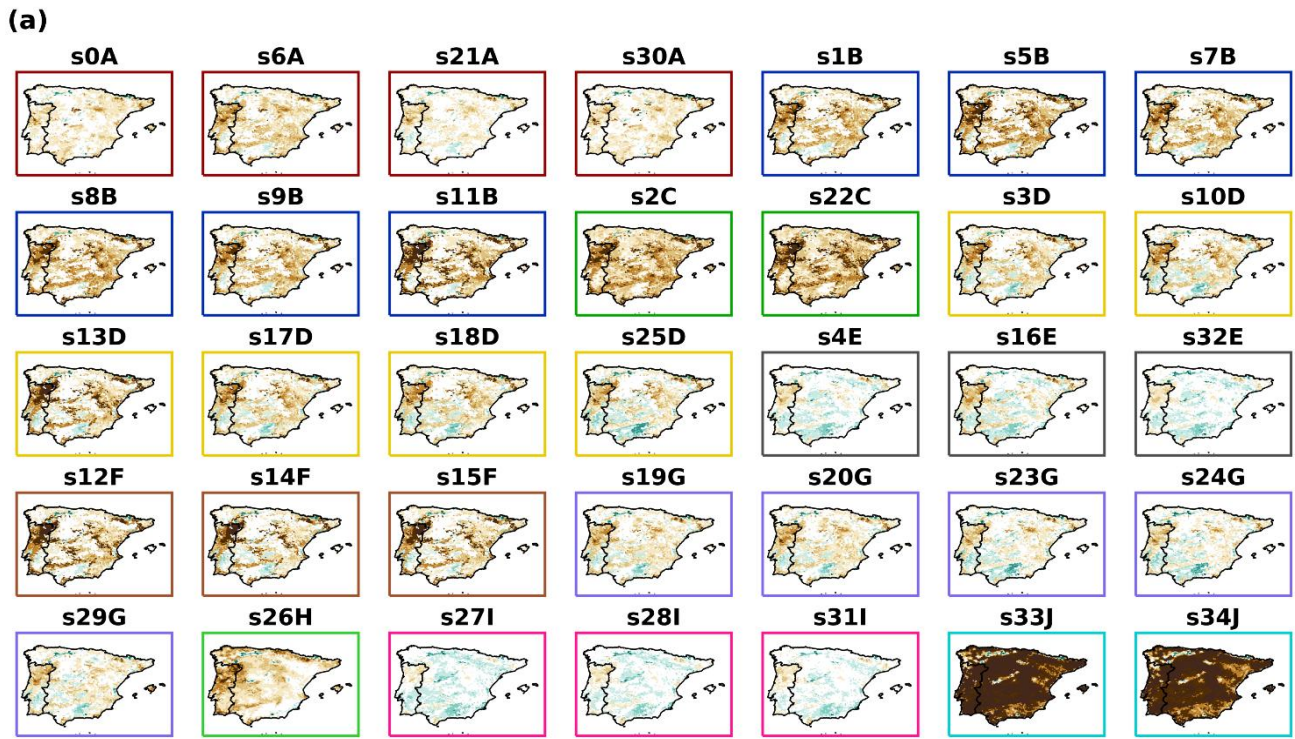
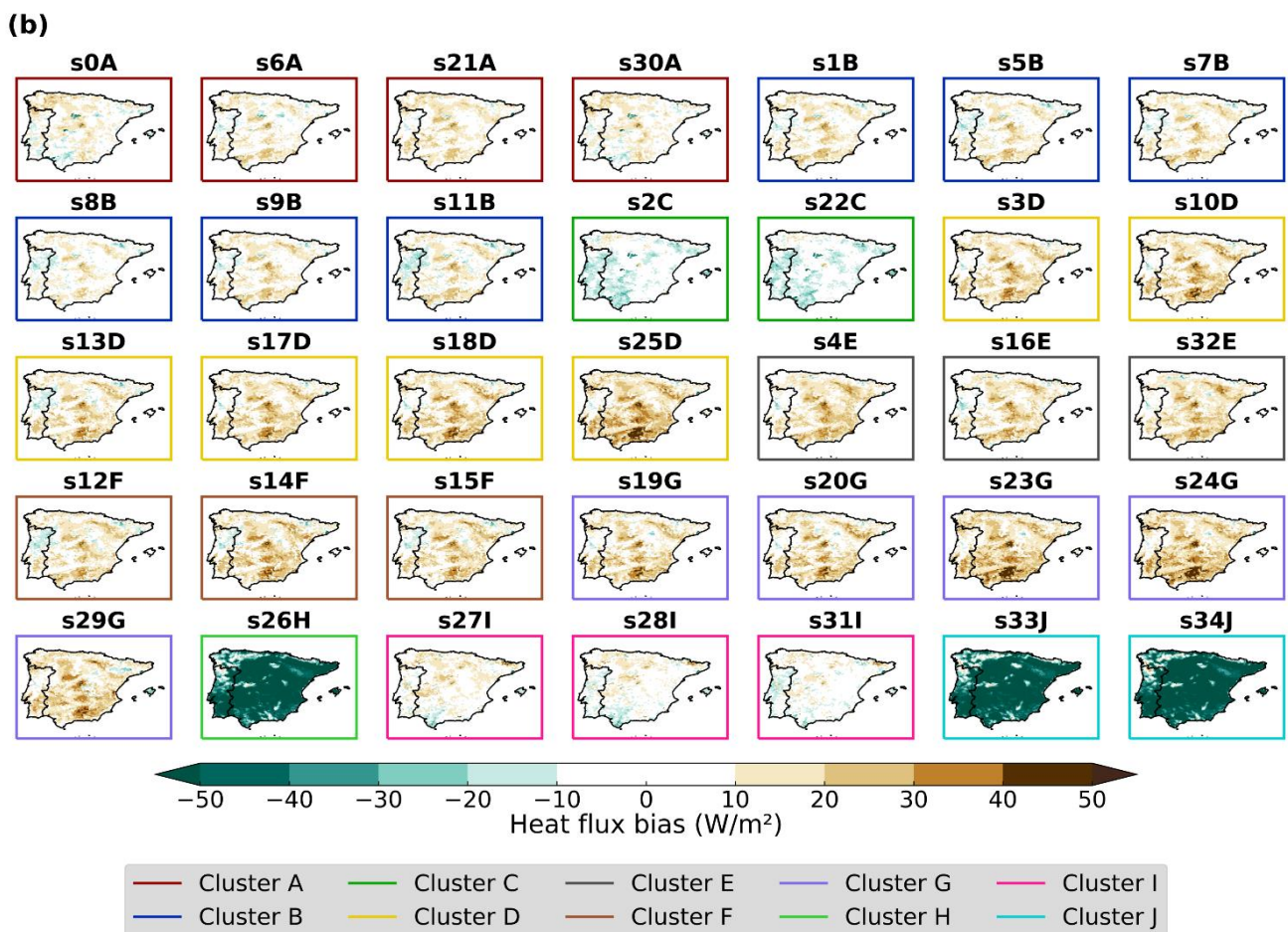

Figure 5

(a)

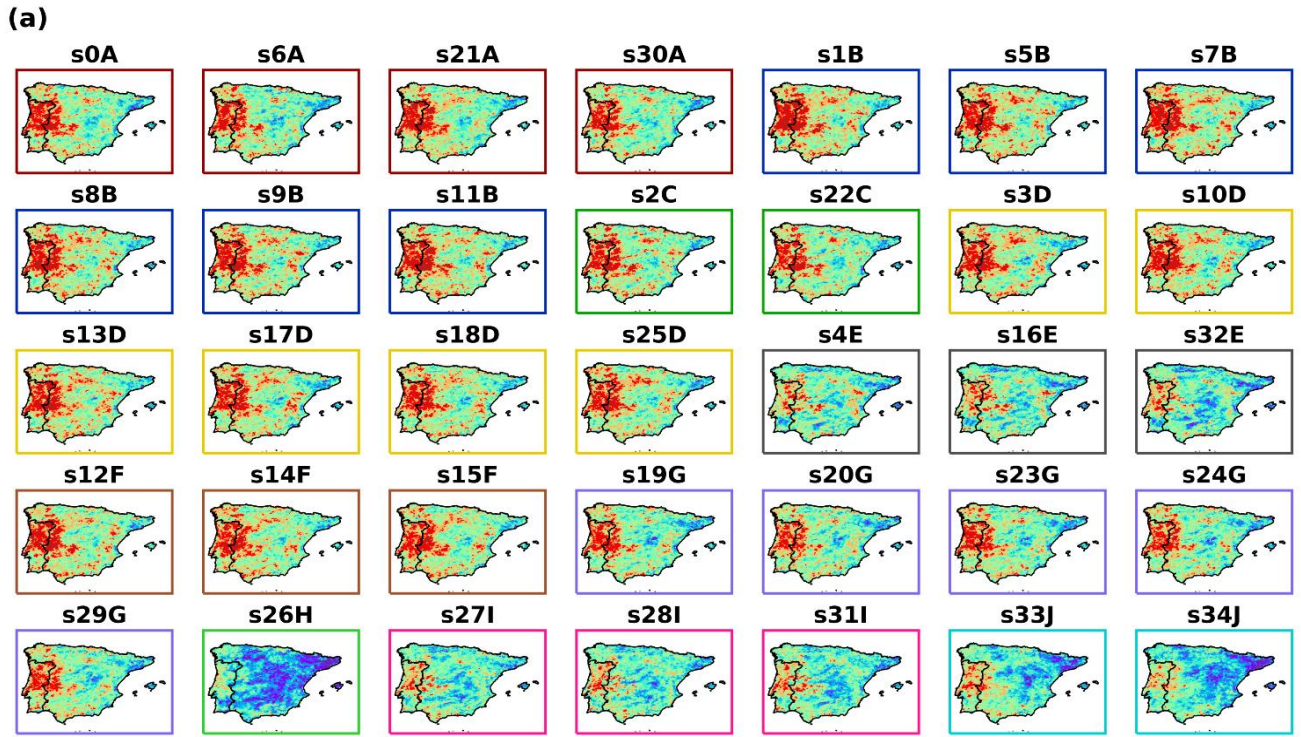

(b)

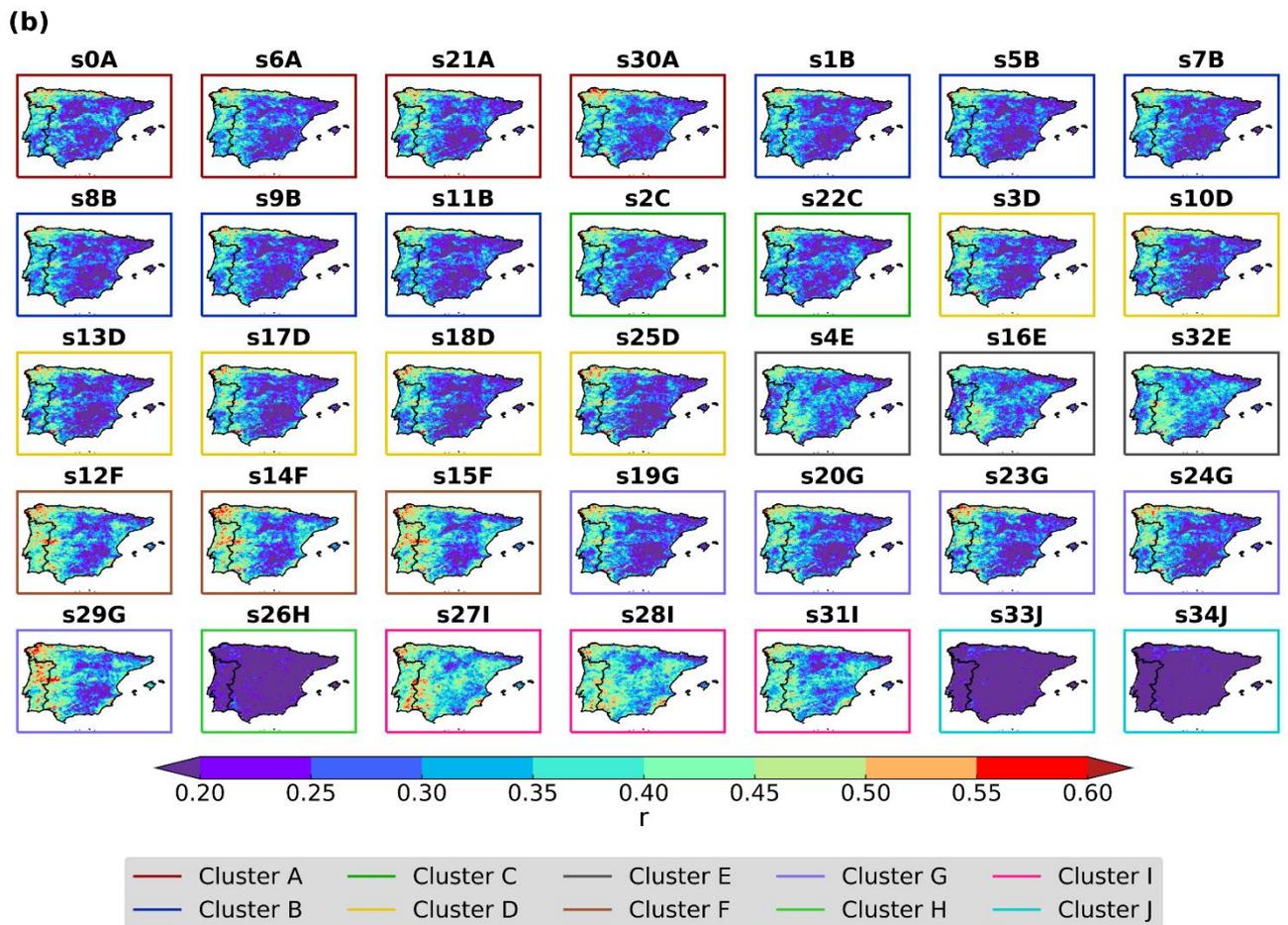

Figure 6

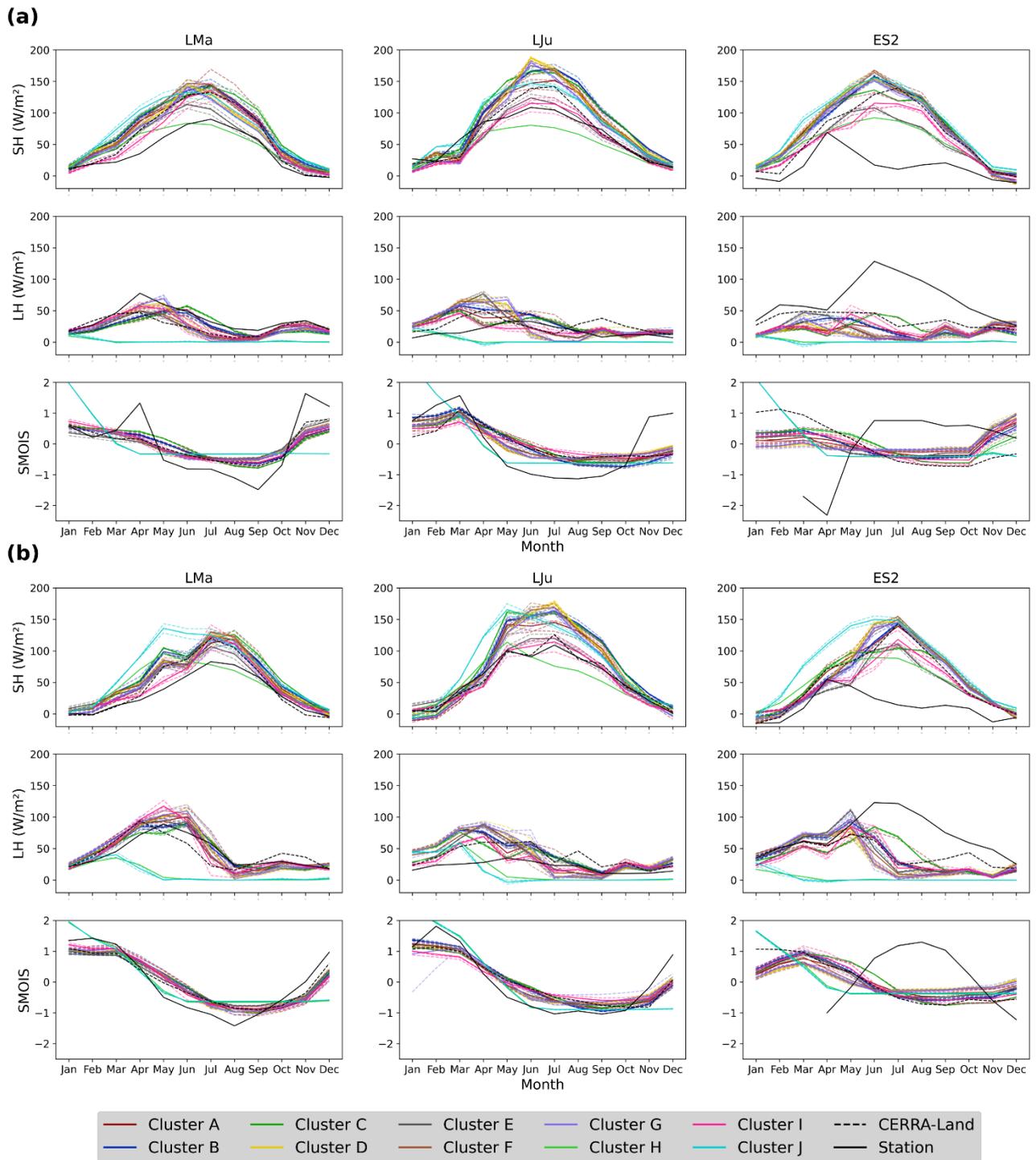

**Figure 7**

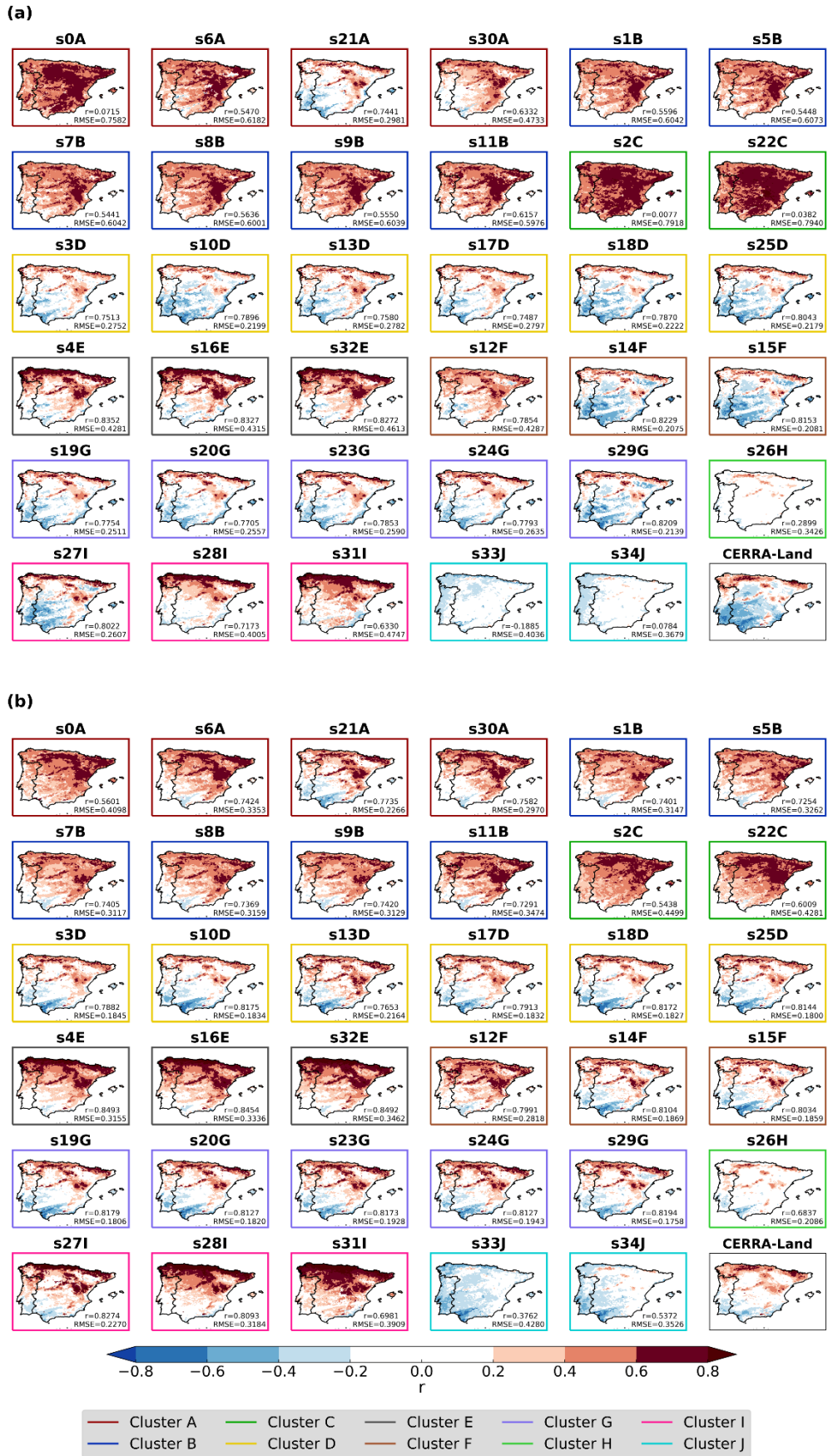

**Figure 8**

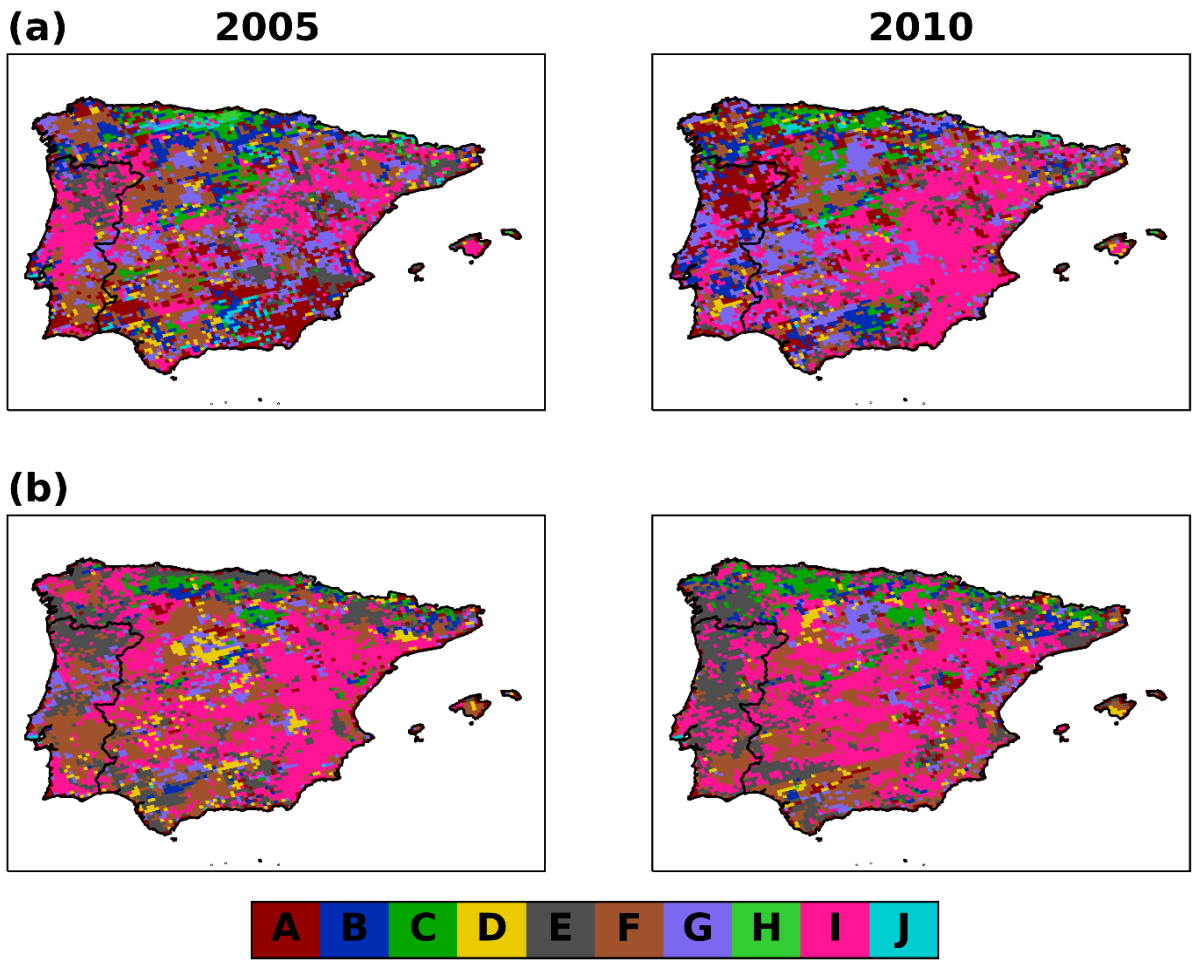

**Figure 9**